\documentclass[aps,pre,nofootinbib,amsmath,amssymb,amsfonts,twocolumn,showpacs]{revtex4-1}
\usepackage{graphicx}
\usepackage{color}
\usepackage{bbm}
\usepackage[utf8]{inputenc}
\usepackage{dcolumn}
\usepackage{hyperref}
\hypersetup{colorlinks=true,breaklinks,linkcolor=blue,urlcolor=blue,citecolor=blue}

\newcommand{\kv}{\ensuremath{p}}
\newcommand{\qv}{\ensuremath{q}} 

\newcommand{\abs}[1]{\ensuremath{\lvert#1\rvert}}
\newcommand{\Ttau}{\ensuremath{T_{\tau}}}
\newcommand{\dtau}{{\ensuremath{\partial_\tau}}}
\renewcommand{\i}{{\ensuremath{i}}}
\newcommand{\iom}{{\ensuremath{\i\omega}}}
\newcommand{\inu}{{\ensuremath{\i\nu}}}
\newcommand{\av}[1]{\ensuremath{\left\langle #1 \right\rangle}}
\newcommand{\um}{\ensuremath{\mathbbm{1}}}
\newcommand\Let{\mathrel{\mathop:\!\!=}}
\newcommand\teL{\mathrel{=\!\!\mathop:}}

\def \sgn {\mathop {\rm sgn}}
\def \Im {\mathop {\rm Im}}
\def \Re {\mathop {\rm Re}}

\begin{document}

\title{Self-energy and vertex functions from hybridization expansion continuous-time quantum Monte Carlo for impurity models with retarded interaction}

\author{Hartmut Hafermann}
\affiliation{Institut de Physique Th\'eorique (IPhT), CEA, CNRS, 91191 Gif-sur-Yvette, France}

\date{\today}

\begin{abstract}
Optimized measurements for the susceptibility, self-energy, as well as three-leg and four-leg vertex functions are introduced for the continuous-time hybridization expansion quantum Monte Carlo solver for the impurity model in the presence of a retarded interaction.
The self-energy and vertex functions are computed from impurity averages which involve time integrals over the retarded interaction. They can be evaluated efficiently within the segment representation.
These quantities are computed within dynamical mean-field theory in the presence of plasmonic screening. In the antiadiabatic regime, the self-energy is strongly renormalized but retains features of the low-energy scale set by the screened interaction. An explicit expression for its high-frequency behavior is provided. 
Across the screening-driven and interaction-driven metal-insulator transitions, the vertex functions are found to exhibit similar structural changes, which are hence identified as generic features of the Mott transition.
\end{abstract}


\maketitle

\section{Introduction}
The continuous-time hybridization expansion quantum Monte Carlo algorithm~\cite{Werner06} (CT-HYB) is an important numerical tool in the context of dynamical mean-field theory (DMFT)~\cite{Georges96}.
Restricting the general two-fermion interaction to density-density terms greatly simplifies the structure of the fermionic trace which enters the Monte Carlo weight. 
An individual Monte Carlo configuration can diagrammatically be depicted in terms of segments indicating intervals of occupancy of the impurity. This ``segment picture'' variant (CT-SEG) of the algorithm provides extremely efficient sampling and measurements for single-site multi-orbital impurity models with density-density interaction.

A further advantage of this method, which has been exploited more recently, is the fact that it can be applied to problems which involve a coupling of the impurity charge to bosons. Using the Lang-Firsov transformation to eliminate the electron-boson coupling leads to a modified Monte Carlo weight which contains an interaction between all pairs of hybridization events and can be computed at essentially no additional computational cost~\cite{Werner07}. The CT-HYB algorithm can further be generalized to treat the coupling of the impurity spin to a vector bosonic field~\cite{Otsuki13}.

Integrating out the bosonic degrees of freedom leads to a retarded interaction among the impurity electrons. By introducing an auxiliary bath of bosonic modes, an arbitrary frequency dependence can be treated without approximation~\cite{Werner10}.
This has been used in realistic simulations of correlated materials (``LDA+DMFT'') to account for the screening effect through a generally complicated frequency dependence of the retarded interaction~\cite{Werner12}.

Another domain in which the method has been applied~\cite{Ayral13}, is the extended dynamical mean-field theory (EDMFT). EDMFT was developed in the context of spin-glasses~\cite{Parcollet99} and strongly correlated electron systems to treat the effect of nonlocal Coulomb interaction~\cite{Smith00,Chitra00}. The screening effect due to the nonlocal interaction leads to a retarded local interaction described by a bosonic bath. 
The latter is determined self-consistently by relating the impurity and lattice susceptibilities.

The solver is further important for the implementation of the recently proposed dual boson approach~\cite{Rubtsov12}, which may be viewed as a diagrammatic extension of EDMFT. Nonlocal corrections to the EMDFT self-energy and polarization are included through a perturbation series whose elements contain the three-leg and four-leg vertices of the impurity problem.
Apart from this, impurity vertex functions are important for the evaluation of momentum resolved response functions within DMFT~\cite{Georges96} and its diagrammatic extensions~\cite{Toschi07,Rubtsov08,Rubtsov12,Rohringer13}.

In the present paper, improved measurements for the susceptibility, self-energy and vertex functions of the impurity model are provided, which are relevant to the above applications. The self-energy and vertex functions are computed from higher-order correlation functions using relations obtained from the equation of motion. The idea was first applied in the numerical renormalization group (NRG) context~\cite{Bulla98,Bulla08}, and has also proven very useful for the CT-HYB algorithm~\cite{Hafermann12}. In the presence of an electron-boson coupling, these improved estimators have to be modified. The coupling to bosons gives rise to additional correlation functions involving the Bose operators~\cite{Hewson02}. Here it is shown that they can solely be expressed in terms of impurity averages, which involve time integrals over the retarded interaction. In the segment representation, these averages can be evaluated without approximation and at small additional computational cost.

The paper is organized as follows:
The Hamiltonian and action formulation of the impurity problem are introduced in Sec.~\ref{sec:impurity}. After briefly reviewing the CT-SEG algorithm in the presence of a retarded interaction in Sec.~\ref{sec:ctseg}, the test case for subsequent calculations is defined (Sec.~\ref{sec:impesttest}). How to improve  the individual measurements is the focus of Sec.~\ref{sec:measurements}. 
Results for the self-energy and vertex functions in the presence of plasmonic screening computed within DMFT on the Bethe lattice are presented and discussed in Sec.~\ref{sec:results}.
Following the conclusions and outlook in Sec.~\ref{sec:conclusions}, a detailed derivation of the relations used in the paper is provided in the Appendices.

\section{Impurity model}
\label{sec:impurity}

We consider a multi-orbital Anderson impurity model with Hubbard-type density-density interaction. Since the technical aspects discussed in this paper are relevant also for realistic calculations involving multiple orbitals, the notation is kept general: The expressions are given for a multi-orbital model with potentially off-diagonal hybridization.

\subsection{Hamiltonian formulation}

The CT-SEG algorithm can be formulated using the action representation of the impurity model. The derivation of the improved estimators from the equation of motion however is based on the Hamiltonian formulation, which we state first.
The Hamiltonian representation has the following form:
\begin{align}
\label{hamiltonian}
H =H_{\text{at}} + H_{\text{bath,F}} + H_{\text{hyb,F}} + H_{\text{bath,B}} + H_{\text{coupling,B}},
\end{align}
where
\begin{align}
H_{\text{at}} =& \sum_i (\varepsilon_i-\mu) n_i+ \frac{1}{2}\sum_{ij} U_{ij} n_i n_j,\\
H_{\text{bath,F}} =& \sum_{\kv i} \tilde{\varepsilon}_{\kv i} f_{\kv i}^\dagger f_{\kv i},\\
H_{\text{hyb,F}} =& \sum_{\kv ij}\left(c_i^\dagger V_{\kv ij} f_{\kv j} + f^\dagger_{\kv i} V_{\kv ij}^{*} c_j \right),\\
\label{hboson1}
H_{\text{bath,B}} =& \sum_{\qv} \omega_{\qv} b_{\qv}^{\dagger}b_{\qv},\\
\label{hboson2}
H_{\text{coupling,B}} =& \sum_{\qv}(b_{\qv}^{\dagger} + b_{\qv})\lambda_{\qv}\sum_{i} n_{i}.
\end{align}
Here latin indices label flavor (spin-orbital) indices of the impurity. $H_{\text{at}}$ is the atomic part of the Hamiltonian, corresponding to a free atom with static density-density interaction $U_{ij}$, energy levels $\varepsilon_{i}$ and chemical potential $\mu$. $H_{\text{bath}}$ describes a bath of noninteracting fermions with dispersion $\tilde{\varepsilon}_{\kv i}$, where $\kv$ labels the fermionic bath states.
The hybridization of the bath electrons with the impurity is mediated by $H_{\text{hyb}}$, where $V_{\kv ij}$ are the hybridization matrices which allow an impurity electron of a given flavor to hybridize with bath electrons of any other. The term $H_{\text{bath,B}}$ describes a bath of free bosons, with bath states labeled by $\qv$. For simplicity, the impurity electrons are assumed to couple to the bosons only through the total charge density of the impurity $n=\sum_{i}n_{i}$, as described by the last term. This restriction can be relaxed in the present algorithm, as described below. $\lambda_{\qv}$ is the coupling strength.

\subsection{Impurity action}

Integrating out the fermionic and bosonic bath degrees of freedom (see Appendix~\ref{app:impurity}) results in an action of the following form:
\begin{align}
\label{action1}
S = S_{\text{at}} + S_{\text{hyb}} + S_{\text{ret}}
\end{align}
where the atomic, hybridization and retarded interaction parts read
\begin{align}
\label{sat}
S_{\text{at}} =& -\!\int_{0}^{\beta}\!\!d\tau\int_{0}^{\beta}\!\!d\tau'\sum_{ij}c_{i}^{*}(\tau)\mathcal{G}^{\text{at}\,-1}_{ij}(\tau-\tau')c_{j}(\tau')\notag\\
 &+\frac{1}{2}\sum_{ij}U_{ij}\int_{0}^{\beta}\!\!\!d\tau\, n_{i}(\tau)n_{j}(\tau),\\
\label{shyb}
S_{\text{hyb}} =& \!\int_{0}^{\beta}\!\!d\tau\int_{0}^{\beta}\!\!d\tau'\sum_{ij}c_{i}^{*}(\tau)\Delta_{ij}(\tau-\tau')c_{j}(\tau'),\\
\label{sret}
S_{\text{ret}} =& \frac{1}{2}\int_{0}^{\beta}d\tau\int_{0}^{\beta}d\tau'\sum_{ij}n_{i}(\tau)U_{\text{ret}}(\tau-\tau')n_{j}(\tau')
\end{align}
and the atomic propagator in \eqref{sat} is given by
\begin{align}
\mathcal{G}^{\text{at}}_{ij}(\inu_{n}) = & (\inu_{n} + \mu - \epsilon_{i})^{-1}\delta_{ij}.
\end{align}
The hybridization function and retarded interaction read in terms of their spectral representations
\begin{align}
\Delta_{ij}(\inu) = & -\int_{-\infty}^{\infty} \frac{d\omega'}{\pi} \frac{\Im \Delta(\omega')}{\inu-\omega'},\\
\label{Uret}
U_{\text{ret}}(\iom) =  & -\int_{0}^{\infty} \frac{d\omega'}{\pi} \Im U_{\text{ret}}(\omega') \frac{2\omega'}{(\iom)^{2}-\omega'^{2}},
\end{align}
where
\begin{align}
-\frac{1}{\pi}\Im\Delta_{ij}(\omega') =  & \sum_{\kv} \sum_{l} V_{\kv il}\delta(\omega'-\tilde{\varepsilon}_{\kv l})V_{\kv lj}^{*} ,\\
-\frac{1}{\pi}\Im U_{\text{ret}}(\omega') = & \sum_{\qv}\lambda_{\qv}\delta(\omega'-\omega_{\qv})\lambda_{\qv}.
\end{align}
The hybridization amplitudes $V_{\kv ij}$ and bath levels $\tilde{\varepsilon_{\kv l}}$, as well as the couplings $\lambda_{\qv}$ and boson frequencies $\omega_{\qv}$, may be chosen to produce a given frequency dependence of the hybridization function and retarded interaction.

In terms of the Boson frequencies and coupling constants, the retarded interaction in \eqref{Uret} is explicitly given by
\begin{align}
U_{\text{ret}}(\iom) = -\sum_{\qv}\frac{2\lambda_{\qv}^{2}\omega_{\qv}}{\omega_{\qv}^{2}-(\i\omega)^{2}}
\end{align}
and the full frequency-dependent interaction of the model reads
\begin{align}
U_{ij}(\iom) = U_{ij} + U_{\text{ret}}(\iom).
\end{align}
In the infinite frequency limit, or for large frequencies compared to a characteristic frequency $\omega_{\qv}$ in case of a single dominant bosonic mode, the interaction $U(\iom)$ approaches the bare interaction: $U(\iom\to\infty)=U+U_{\text{ret}}(\iom\to\infty)=U$. This expresses the fact that screening becomes less effective at high energies.
On the other hand, in the static limit and for small frequencies compared to the characteristic frequency, the interaction is given by a smaller screened value
\begin{align}
\label{uscr}
U_{ij}^{\text{scr}}\equiv U_{ij}+U_{\text{ret}}(\iom=0) = U_{ij} - \sum_{\qv}\frac{2\lambda_{\qv}^{2}}{\omega_{\qv}} < U.
\end{align}

\section{CT-SEG algorithm}
\label{sec:ctseg}

\begin{figure}[t]
\begin{center}
\includegraphics[scale=0.45,angle=0]{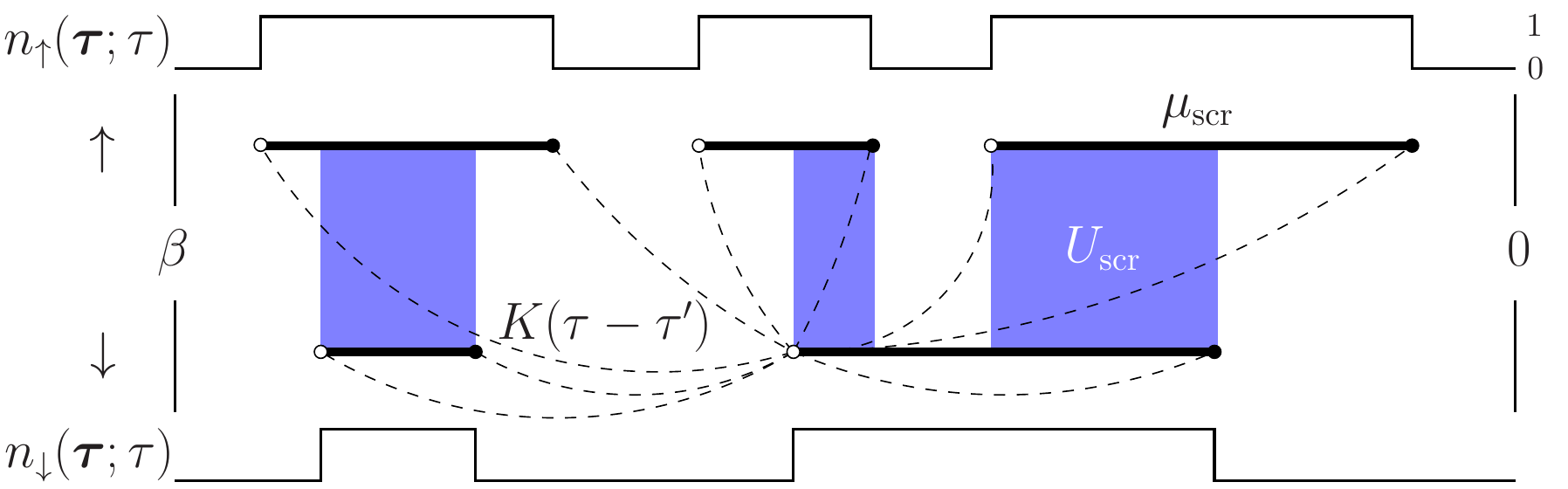} 
\end{center}
\caption{\label{fig:ctseg} (Color online) Illustration of a configuration in the CT-SEG algorithm. There is a time line for each flavor (spin) and segments mark the time intervals where the impurity is occupied. Creation operators are marked by closed circles and annihilators by open ones. The realization of the density is piecewise constant and changes between 0 (unoccupied) and 1 (occupied) at the operator positions. Blue shaded areas indicate the time-intervals for which the impurity is doubly occupied. The retarded interaction kernel $K(\tau-\tau')$ connects a given operator with all other operators in the configuration, as indicated by the dashed lines.
}
\end{figure}

The basic idea underlying the CT-SEG algorithm~\cite{Werner06} is to expand the partition function in the impurity-path hybridization term, $Z = Z_{\text{at}}\sum_{k=0}^{\infty} \int d\boldsymbol{\tau}\, w(\boldsymbol{\tau})$, where $k$ denotes the perturbation order. Here $\boldsymbol{\tau}=(\tau_{2k},\ldots,\tau_{1})$ is a time-ordered sequence of imaginary time points which specify the operator positions where the hybridization events occur. In order to fully specify a Monte Carlo configuration, which we symbolically denote by $\boldsymbol{\tau}$, we additionally need to keep track of which flavor a given time corresponds to. A configuration may then be depicted by a number of ``segments'' for each flavor, representing the time intervals during which the impurity is occupied and hence the name of the algorithm.
Assuming $w(\boldsymbol{\tau})$ is positive (otherwise we rewrite $w(\boldsymbol{\tau})=\sgn[w(\boldsymbol{\tau})]\abs{w(\boldsymbol{\tau})}\,$), we can sample configurations using a Metropolis algorithm. The Monte Carlo average of an observable $O$ is given by $\av{O}_{\text{MC}}\Let \sum_{k=0}^{\infty} \int d\boldsymbol{\tau}\, O(\boldsymbol{\tau})w(\boldsymbol{\tau})/\sum_{k=0}^{\infty} \int d\boldsymbol{\tau}\,w(\boldsymbol{\tau})$, where $O(\boldsymbol{\tau})$ denotes the realization of the observable $O$ in a given configuration.
The Monte Carlo weight is the product $w(\boldsymbol{\tau}) =  w_{\text{hyb}}(\boldsymbol{\tau})w_{\text{at}}(\boldsymbol{\tau}) w_{\text{ret}}(\boldsymbol{\tau})$, where $w_{\text{hyb}}(\boldsymbol{\tau})$ can be expressed as a determinant of hybridization functions.
The remaining weights are computed as a trace  over the atomic states.
For a given configuration (see Fig.~\ref{fig:ctseg}), the realization of the density is piecewise constant and changes at the operator positions (kinks). Exploiting this fact makes the algorithm particularly efficient. For example, the evaluation of the atomic part of the weight simply amounts to counting the length of the segments $l_{ii}$ and their overlap $l_{ij}$: $w_{\text{at}}(\boldsymbol{\tau})=e^{-\frac{1}{2}\sum_{ij}U_{ij}l_{ij}}e^{\mu\sum_{i}l_{ii}}$. For the retarded part one has to evaluate the exponential of the double integral in \eqref{sret} for a given configuration. To this end, it is convenient to define the retarded interaction kernel $K(\tau)$ such that its second derivative yields the retarded interaction\footnote{While simply being defined here as the twice-integrated retarded interaction, $K(\tau)$ emerges naturally when performing an expansion in the electron phonon coupling $H_{\text{coupling}}$ and evaluating the thermal average over products of phonon fields~\cite{Otsuki13}.}. 
$K(\tau)$ and its derivatives obey bosonic symmetry, i.e. $K(-\tau)=K(\beta-\tau)$. $K(\tau)$ and $U(\tau)$ are also symmetric around zero, i.e. $K(-\tau)=K(\tau)$, while its derivative $K'(\tau)$ is antisymmetric around zero. $K(\tau)$ correspondingly has a slope discontinuity at $\tau=0$.
Taking this into account and imposing the boundary conditions $K(\beta)=K(0^{+})=0$, the weight due to the retarded interaction is~\cite{Werner10}
\begin{align}
w_{\text{ret}}(\boldsymbol{\tau}) =& e^{\frac{1}{2}\sum_{ij}\sum_{2k_{i,j}\geq \alpha_{i},\alpha_{j}> 0} ^{\alpha_{i}\neq\alpha_{j}}s_{\alpha_{i}}s_{\alpha_{j}}K(\tau_{\alpha_{i}}-\tau_{\alpha_{j}})}\notag\\
&\times e^{ 2K'(0^{+})\sum_{ij}^{i\neq j}l_{ij} + K'(0^{+})\sum_{i}l_{ii}},
\end{align}
where the second sum in the first line is over the $2k_{i}$ operators at positions $\tau_{\alpha_{i}}$ in channel $i$. The sign $s_{\alpha}=\pm 1$ is positive if the operator at time $\tau_{\alpha}$ is a creator and negative for an annihilator. 
The second line has the same form as the weight $w_{\text{at}}$. With $U_{ij}^{\text{scr}}= U_{ij} + U_{\text{ret}}(\iom=0)$ and $U_{\text{ret}}(\iom=0)=\int_{0}^{\beta}d\tau U_{\text{ret}}(\tau)=-2K'(0^{+})$, one therefore finds that the chemical potential and static interaction have to be shifted internally in the solver according to
\begin{align}
\label{musolver}
\mu &\to\mu^{\text{solver}} = \mu + \frac{U_{ij}-U_{ij}^{\text{scr}}}{2} = \mu + K'(0^{+}),\\
\label{Usolver}
U_{ij} &\to U_{ij}^{\text{solver}} =  U_{ij}^{\text{scr}} = U_{ij}-2K'(0^{+}).
\end{align}

The electron-boson coupling \eqref{hboson2}  through the full density is sufficiently general for a number of applications involving bosons: This includes phonons or the problem of screening through a retarded interaction with a given frequency dependence.
A spin-boson coupling of the form $\sum_{\qv}(b_{\qv}^{\dagger} + b_{\qv})\lambda_{\qv}S_{z}$ requires us to differentiate between the couplings to individual densities $n_{i}$. This case can be treated within the present algorithm and the generalization is straightforward\footnote{In this case, the boson operators acquire a flavor index and a bosonic bath is introduced for each flavor, i.e. $H_{\text{bath,B}}=\sum_{\qv i}\omega_{\qv i}b_{\qv i}^{\dagger},b_{\qv i}$.
The coupling constants become matrices $\lambda_{\qv ij}$ and the spectral density is given by $-(1/\pi)\Im \mathcal{U}(\omega)=\sum_{\qv}\sum_{l}\lambda_{\qv il}\delta(\omega-\omega_{\qv l})\lambda_{\qv lj}$. Final expressions (e.g. for the improved estimators) are generalized by simply attaching orbital indices to the retarded interaction in complete analogy to the static case.}.
The treatment of the coupling of the impurity spin to a vector bosonic field requires a secondary expansion in the bosonic bath in addition to the expansion in the hybridization~\cite{Otsuki13} and is beyond the scope of the present work.

\section{Test case}
\label{sec:impesttest}

In order to test the improved estimators and to study the effect of the retarded interaction on the self-energy and vertex functions, we will use the following test case throughout the paper: We consider the Hubbard model with static interaction $U$ and in the presence of plasmonic screening within DMFT on the Bethe lattice. The full bandwidth is $W/t=4$ and the temperature is fixed at $T/t=0.02$. Energies will be measured in units of the hopping $t$. For the model with static interaction only, the Mott transition occurs at $U/t\approx 5.1$.

It is instructive to consider a retarded interaction originating from a single plasmon (or phonon) mode with a characteristic screening frequency $\omega_{0}$: $-\frac{1}{\pi}\Im U_{\text{ret}}(\omega)=\lambda^{2}\delta(\omega-\omega_{0})$.
With the definition $K''(\tau)=U_{\text{ret}}(\tau)$ and boundary conditions $K(0)=K(\beta)=0$, the retarded interaction kernel and its derivative for this case are given by the expressions [see Eqs. \eqref{dtau} and \eqref{uretdef}]:
\begin{align}
K(\tau) =& -\frac{\lambda^{2}}{\omega_{0}^{2}} \frac{\cosh[\omega_{0}(\tau-\beta/2)]-\cosh(\omega_{0}\beta/2)}{\sinh(\omega_{0}\beta/2)},\\
K'(\tau) =& \frac{\lambda^{2}}{\omega_{0}} \frac{\sinh[\omega_{0}(\tau-\beta/2)]}{\sinh(\omega_{0}\beta/2)}.
\end{align}

The case where the static bare (unscreened) interaction is large and the screened interaction is significantly smaller is particularly interesting. In the following, the static interaction is kept fixed to $U=8$, which is equal to twice the bandwidth. Without screening, the system would hence be insulating. The screening frequency $\omega_{0}$ is varied while choosing the electron-boson coupling $\lambda=\sqrt{(U-U_{\text{scr}})\omega_{0}/2}$ such that the screened interaction is fixed to $U_{\text{scr}}=3<W$. The parameters are the same as in Ref.~\onlinecite{Werner10}.

\section{Measurements}
\label{sec:measurements}

Because of the simple structure of the trace, the evaluation of the Monte Carlo weight is very efficient. The performance of individual measurements is therefore critical for the overall performance of the algorithm.
The efficiency of the implementation can be improved by exploiting the structure of the trace and/or by better utilizing the information that is available, i.e. by introducing so-called improved estimators. These are additional correlation functions which combine with Green's functions through expressions that follow from the equation of motion, in order to give more accurate estimates for the self-energy and vertex functions.

\subsection{Charge susceptibility}

An illustrative example of how the structure of the trace may efficiently be exploited to improve performance are susceptibilities which can be written in terms of averages over products of density operators. This applies to the important cases of the spin- and charge susceptibilities. For simplicity, we restrict ourselves to the latter. For other types of susceptibilities, such as the pairing susceptibility $\chi_{pp}(\tau-\tau') =\langle c^{\phantom{\dagger}}_{\uparrow}(\tau)c^{\phantom{\dagger}}_{\downarrow}(\tau)c^{\dagger}_{\downarrow}(\tau')c^{\dagger}_{\uparrow}(\tau')\rangle$, a similar approach unfortunately does not exist.
Without the improvements discussed here, the susceptibility measurement can become the bottleneck of the calculation and may be difficult to converge in practice. 

The impurity charge susceptibility
\begin{align}
\label{chi}
\chi(\iom) = -\sum_{ij}\Big(\av{n_{i}(\iom)n_{j}(-\iom)} -\av{n_{i}}\av{n_{j}}\delta_{\omega}\Big)
\end{align}
is determined from the density-density correlation function $\chi_{ij}(\tau-\tau')\equiv-\av{n_{i}(\tau)n_{j}(\tau')}$. This measurement differs essentially from the one for Green's function. The latter is usually measured as a ratio of determinants:\footnote{It may, in principle, be measured as a ratio of traces, which in general is not ergodic: Nonzero contributions to Green's function exist which can only be obtained by inserting operators into a configuration which has a vanishing weight and hence is never sampled.} When measured in imaginary time, it is binned on a fine grid and measured at $k^{2}$ time differences, where $k$ is the current perturbation order. Hence the algorithm does not scale with the grid size which can (and should) be chosen large.

The density-density correlation function, on the other hand, is most simply measured as a ratio of time-ordered traces\footnote{It can also be measured using the shift operator method~\cite{Augustinsky13}.}. In the segment picture, the ratio of the traces with an operator $n(\tau)$ inserted at time $\tau$ to the trace without this operator is simply $1$ if a segment is present (i.e. $\tau$ follows a creation operator) and 0 otherwise.
The density for a particular realization of a Monte Carlo configuration defined by $\boldsymbol{\tau}$ is hence a piecewise constant function which changes at the operator positions (kinks), as illustrated in Fig.~\ref{fig:ctseg}. We denote it by $\tilde{n}_{i}(\boldsymbol{\tau};\tau)$.

In a time measurement, the contribution to the susceptibility is computed in two steps. First, $\tilde{n}(\boldsymbol{\tau};\tau)$ is computed on a grid. This operation is linear in the number of imaginary time bins and independent of the perturbation order $k$. A measurement for $\chi_{ij}(\tau-\tau')$ within a given Monte Carlo configuration $\boldsymbol{\tau}$ is given by the product $\tilde{n}_{i}(\boldsymbol{\tau};\tau)\tilde{n}_{j}(\boldsymbol{\tau};\tau')$, which has to be evaluated for all time differences. The complexity of the measurement is hence dominated by the second step, which scales \emph{quadratically} with the bin number $N_{\tau}$. Note that in general, the latter is significantly larger than $k$.

\begin{figure}[t]
\begin{center}
\includegraphics[scale=0.7,angle=0]{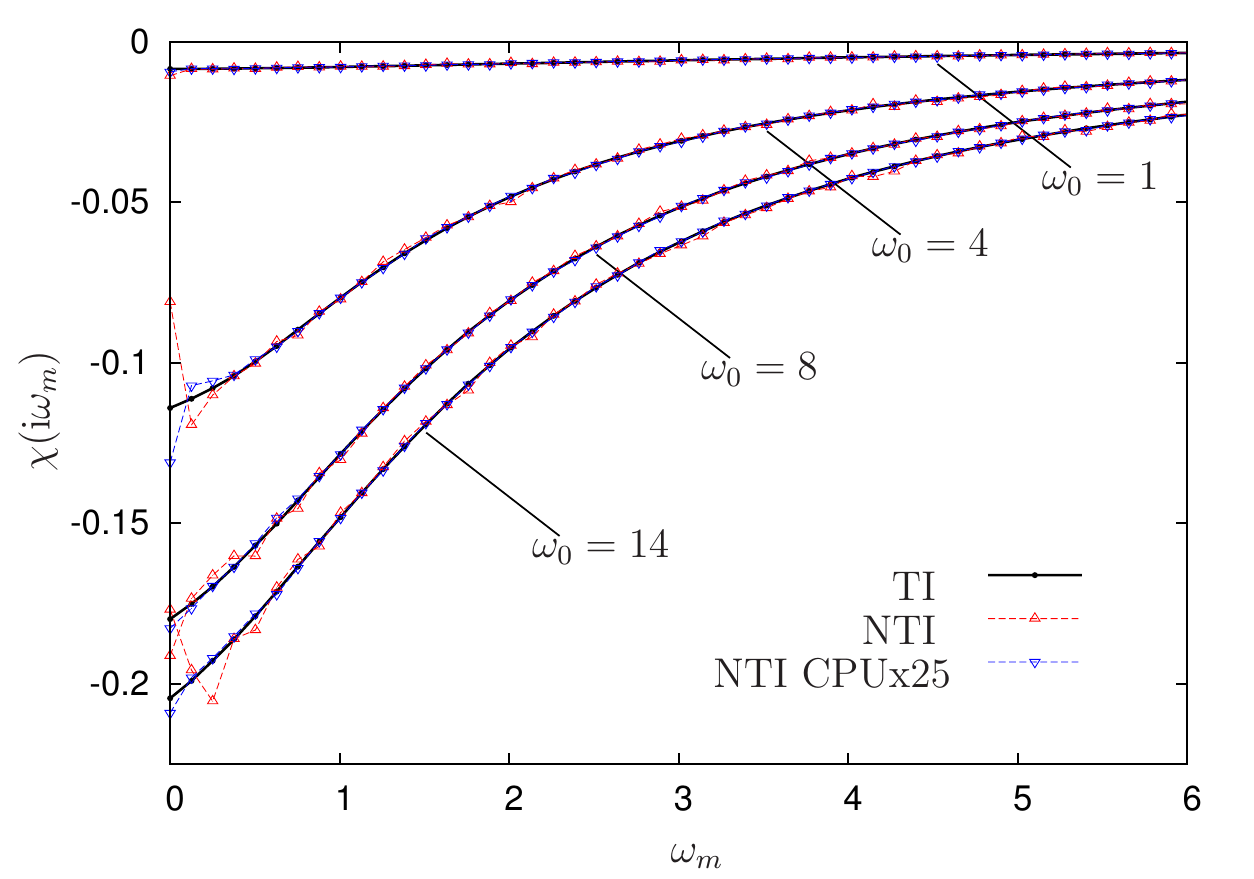} 
\end{center}
\caption{\label{fig:chiw} (Color online) Impurity charge susceptibility $\chi(\iom)$ for different screening frequencies $\omega_{0}$.
Results are shown for the translationally invariant (TI) measurement (heavy solid lines, closed symbols) and without using time translational invariance (NTI, thin dashed lines, open symbols), for a runtime of 300s on 16 cores.
While the TI results are converged (for these results the error bar is smaller than the symbol size), the NTI results exhibit large errors in particular at low frequencies. Results for the NTI measurement with the runtime increased by a factor of $N=25$ still exhibit substantial errors at low frequencies, illustrating the slow convergence. All results are consistent with a decrease in error bar by a factor of $\sqrt{N}$, except for the NTI measurement at $\omega_{0}=4$, relatively close to the Mott transition, showing that the NTI measurement is particularly sensitive to autocorrelation.}
\end{figure}

The measurement is significantly more efficient in frequency. Since $\tilde{n}_{i}(\boldsymbol{\tau};\tau)$ is piecewise constant, we can simply evaluate its Fourier transform with respect to $\tau$. Viewing the imaginary-time interval as a circle\footnote{Using anti-periodic boundary conditions, segments are allowed to overlap from $\tau<\beta$ to $\tau'>0$.}, one obtains
\begin{align}
\label{nw1}
\tilde{n}_{i}(\boldsymbol{\tau};\iom=0) = \sum_{\alpha_{i}=1}^{2k_{i}} s_{\alpha_{i}} \tau_{\alpha_{i}},
\end{align}
which is equal to the total length of the segments (i.e., the occupation) of flavor $i$  and
\begin{align}
\label{nw2}
\tilde{n}_{i}(\boldsymbol{\tau};\iom>0) = \frac{1}{\iom} \sum_{\alpha_{i}=1}^{2k_{i}} s_{\alpha_{i}} \exp(\iom\tau_{\alpha_{i}})
\end{align}
otherwise, with $s_{\alpha_{i}}$ as defined in Sec.~\ref{sec:ctseg}. The exponential for a given frequency should be computed from the previous value as $\exp(\iom_{m+1}\tau) = \exp(\iom_{m}\tau)\exp(\iom_{1}\tau)$ in order to save explicit evaluation of the exponential for all frequencies, which is computationally costly. From the Fourier transform of the density, the measurement of the density-density correlation function is evaluated in a second step as
\begin{align}
\label{nnwmeas}
\chi_{ij}(\iom) = -\av{\tilde{n}_{i}(\boldsymbol{\tau};\iom)\tilde{n}^{*}_{j}(\boldsymbol{\tau};\iom)}_{\text{MC}},
\end{align}
where '$*$' denotes complex conjugate and $\av{\ldots}_{\text{MC}}$ denotes Monte Carlo average over configurations $\boldsymbol{\tau}$ (see Sec.~\ref{sec:ctseg}). While the individual measurements factorize, this is of course no longer true in general for the Monte Carlo average.
The second step scales linearly in the number of frequencies $N_{\omega}$. 
This measurement is hence dominated by the first step which scales as $k N_{\omega}\ll N_{\tau}^{2}$.
From \eqref{nnwmeas}, the charge (and spin) susceptibility \eqref{chi} is computed after the simulation.

The frequency measurement is significantly faster and should be preferred over the imaginary-time measurement. Note that the imaginary time and frequency measurements are equivalent in the sense that they encode, as long as the same configurations are sampled, the \emph{same information} albeit in a different basis (the two operations, the Fourier transform and the Monte Carlo sampling commute). 
For finite resolution in imaginary time, the Nyquist theorem ensures that a function binned on $N_{\tau}$ grid points will reconstruct the function in the frequency domain up to the Nyquist frequency $1/(2\Delta\tau) = N_{\tau}/(2\beta)$. Therefore, to compute the function at $N_{\omega}$ frequencies from the time measurement, one needs a grid size of at least $N_{\tau}\approx4\pi N_{\omega}$, so the speedup of using the frequency measurement is considerable.

Finally, note that measuring $-\av{n_{i}(\tau)n_{j}(0)}$, i.e., not exploiting time translational invariance, effectively does not speed up the calculation. This is illustrated in Fig. \ref{fig:chiw}, where the equivalent frequency measurement $-\av{n_{i}(\iom)n_{j}(\tau=0)}$  is plotted together with the results obtained using \eqref{nnwmeas}, both measured in the same simulation. Because the former does not use time-translational invariance, it converges much more slowly than the latter at low frequencies.

\subsection{Fermionic Self-energy}
\label{sec:impestsigma}

In the CT-HYB algorithm, the extraction of the self-energy from Dyson's equation leads to large numerical errors at intermediate to high frequencies. A similar problem appears for the vertex function. 
The origin of the numerical problems in the evaluation of Dyson's equation is twofold. Firstly, by forming the difference between two functions, their absolute error propagates. Since both Green's functions have numerical errors from different sources\footnote{The noninteracting Green's function may or may not (as in DMFT calculations) be known up to machine precision.}, one cannot expect these errors to cancel. Secondly, the Green's function decays as $1/\inu$, so that the absolute error of the inverses increases rapidly. In interaction expansion continuous-time quantum Monte Carlo (CT-INT) the problem does not exist. The reason is that in CT-INT the Green's function is measured as a correction to the noninteracting Green's function $G_{0}$~\cite{Rubtsov05} (omitting indices):
\begin{align}
\label{ctintmeas}
G = G_{0} + G_{0}\av{M}_{\text{MC}}G_{0},
\end{align}
where $M$ is the inverse of the matrix of noninteracting Green's functions. The measured correction decays at least as $1/(\inu)^{2}$.
Comparing Eq. \eqref{ctintmeas} to Dyson's equation, one sees that $\av{M}_{\text{MC}}G_{0}=\Sigma G$ so that the CT-INT provides direct access to the product $\Sigma G$. The self-energy can be determined from a \emph{ratio} of observables, i.e.
\begin{align}
\label{sigmactint}
\Sigma = \frac{\av{MG_{0}}_{\text{MC}}}{G}.
\end{align}
Note that it is because of \eqref{ctintmeas} that in CT-INT this is equivalent to determining it directly from Dyson's equation\footnote{This is true as far as the Monte Carlo error in $\av{M}_{\text{MC}}$ is concerned: The error propagation is the same because the Taylor expansions of both expressions in $\av{M}_{\text{MC}}$ are identical.},  in contrast to CT-HYB.

In general, for any method which yields Green's functions afflicted with numerical errors, the self-energy should always be determined as a ratio.
However, due to the expansion in the impurity-bath hybridization, it is less obvious how to measure the product $\Sigma G$ in CT-HYB.
The solution is to express $\Sigma G$ in terms of a higher-order correlation function which follows from the equation of motion for Green's function. This technique was first applied successfully in NRG~\cite{Bulla98,Bulla08}. 
In a previous publication~\cite{Hafermann12}, it has been shown that computing the self-energy using this improved estimator also proves very useful for the CT-HYB as it yields substantially more accurate results than the naive approach using Dyson's equation. It should therefore be the method of choice for the determination of the self-energy in CT-HYB.

In the present work, the improved estimators for the self-energy and vertex functions are generalized for the impurity model with retarded interaction. It is important that the resulting expressions can be written solely in terms of impurity averages. They can hence be evaluated without approximation. Despite the retarded character of the interaction, the resulting correlation functions can be evaluated efficiently within the segment representation.

In the following, it will be convenient to switch between the representation of time-ordered averages in terms of operators (which are denoted $c,c^{\dagger}$) and the path integral representation, where the Grassmann numbers are denoted as $c,c^{*}$. For the latter, time-ordering is not explicitly indicated as it is implicit in the construction of the path integral.

\begin{figure}[t]
\begin{center}
\includegraphics[scale=1.0,angle=0]{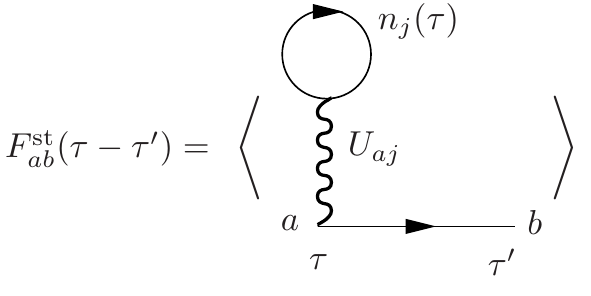} 
\end{center}
\caption{\label{fig:sigmag} Illustration of the static part of the improved estimator $F=\Sigma G$, Eqn. \eqref{fst}. The measurement of this correlation function corresponds to accumulating Hartree-like contributions to the product $\Sigma G$. }
\end{figure}

For a static interaction, the self-energy can be expressed in matrix form as~\cite{Bulla98,Hafermann12}
\begin{align}
\label{sigma}
\Sigma(\inu) = F(\inu)G^{-1}(\inu).
\end{align}
The equation for $F=\Sigma G$ is obtained by considering the equation of motion for Green's function:
\begin{align}
\label{eom}
\dtau G_{ab}(\tau-\tau') = -\delta(\tau-\tau')\delta_{ab} - \av{T_{\tau}[\dtau c_{a}(\tau)]c_{b}^{\dagger}(\tau')},
\end{align}
which involves the commutator with the Hamiltonian, $\dtau c_{a}(\tau) = [H,c_{a}](\tau)$. The time-ordered thermal average of operators $\av{\Ttau\ldots}$ gives rise to the correlation function $F$.
In the case of a static interaction, $F$ essentially stems from the commutator of $c_{a}$ with the interaction term $\sum_{ij}U_{ij}n_{i}n_{j}$.
Switching to the path integral representation, the resulting correlation function is given by~\cite{Hafermann12}
\begin{align}
\label{fst}
F^{\text{st}}_{ab}(\tau-\tau') = -\sum_{j} \av{n_{j}(\tau) U_{ja} c_{a}(\tau)c_{b}^{*}(\tau')}.
\end{align}
The equation of motion obtained by taking the derivative with respect to $\tau'$ generates the corresponding equation for $F'\Let G\Sigma$ (which is the same as $F$ for a diagonal basis).
Diagrammatically, the correlation function \eqref{fst} has the interpretation illustrated in Fig. \ref{fig:sigmag}: When accumulated in the Monte Carlo process, one essentially samples Hartree-like contributions to $\Sigma G$.

\begin{figure}[b]
\begin{center}
\includegraphics[scale=1.0,angle=0]{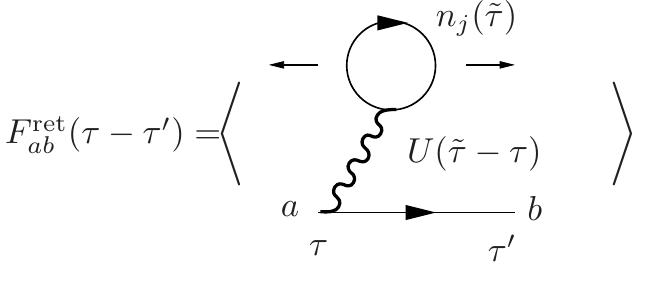} 
\end{center}
\caption{\label{fig:sigmagret} Illustration of the retarded part of the improved estimator, Eqn. \eqref{fret}. As in the static case, one samples Hartree-like contributions to $\Sigma G$. Because the interaction is retarded, one has to integrate over all times $\tilde{\tau}$, as indicated by the arrows.}
\end{figure}

In the dynamic case,  the bath of auxiliary bosons couples to the charge density $n$ through $\sum_{\qv}(b_{\qv}^{\dagger} + b_{\qv})\lambda_{\qv}\sum_{i} n_{i}$. Hence the commutator of $H$ with $c_{a}$ generates an additional term which gives rise to the correlation function
\begin{align}
\label{fret}
F^{\text{ret}}_{ab}(\tau-\tau') = -\sum_{\qv}\lambda_{\qv}\av{T_{\tau} [b_{\qv}^{\dagger}(\tau) + b_{\qv}(\tau)] c_{a}(\tau)c_{b}^{\dagger}(\tau')}.
\end{align}
This result has been obtained previously in the context of NRG for the Anderson-Holstein impurity model. Since NRG is a Hamiltonian based method, the matrix elements involving boson operators have to be computed explicitly~\cite{Hewson02}. Such a treatment involves a truncation of the infinite boson Hilbert space.
In the present algorithm, the improved estimator can be evaluated without approximation: Since the bosonic bath is noninteracting, it is possible to express the correlation function solely in terms of impurity averages. In Appendix~\ref{app:impest} it is shown that integrating out the bosonic bath from \eqref{fret} leads to the expression
\begin{align}
\label{fretfinal}
F^{\text{ret}}_{ab}(\tau-\tau')\! =\! -\!\!\int_{0}^{\beta}\!\!\!d\tilde{\tau} \sum_{i}\av{n_{i}(\tilde{\tau}) U_{\text{ret}}(\tilde{\tau}-\tau) c_{a}(\tau)c_{b}^{*}(\tau')}\!.
\end{align}
This form could have been anticipated from the static result \eqref{fst}: As illustrated in Fig. \ref{fig:sigmagret}, the contributions are also Hartree-like. Due to the retarded nature of the interaction, $U_{\text{ret}}$ has to be integrated over all time differences $\tilde{\tau}-\tau$.
The impurity self-energy in the presence of a retarded interaction is still given by  \eqref{sigma}, where now
\begin{align}
\label{Ftot}
F(\inu) = F^{\text{st}}(\inu) + F^{\text{ret}}(\inu)
\end{align}
is the sum of the static and retarded contributions.
In order to see how the improved estimators are measured, first consider the measurement for Green's function. When measured as as a ratio of determinants of hybridization functions, $G_{ab}(\tau-\tau')\Let-\av{c_{a}(\tau)c^{*}_{b}(\tau')}$ is obtained as follows~\cite{Werner06}:
\begin{align}
\label{gmeas}
G_{ab}(\tau) = -\frac{1}{\beta}\Bigg\langle\sum_{\alpha,\beta=1}^{k}
M_{\beta\alpha}\delta^{-}(\tau,\tau_{\alpha}^{e}-\tau_{\beta}^{s})
\delta_{a,\lambda{\alpha}}\delta_{b,\lambda'_{\beta}}
\Bigg\rangle_{\text{MC}}.
\end{align}
Here $M_{\alpha\beta}$ is an element of the inverse of the matrix of hybridization functions for a given configuration, $k$ is the current perturbation order, $\tau^{s}_{\beta}$ ($\tau^{e}_{\alpha}$) are the times associated with the creators (annihilators) and mark the segment start (end) times, and $\lambda_{\alpha}$ denotes the spin-orbital index associated with the matrix index $\alpha$. $\delta^{-}(\tau,\tau')\Let\sgn(\tau')\delta(\tau-\tau'-\theta(-\tau')\beta)$ is an antisymmetrized $\delta$-function, which transforms a measurement with a negative time difference $\tau_{\alpha}^{e}-\tau_{\beta}^{s}<0$ to one with a positive time according to the identity $G(-\tau)=-G(\beta-\tau)$.
The density is formally measured from a ratio of traces, as discussed before. The static part of the improved estimator can be measured as a combination of the two, i.e.,
\begin{align}
\label{fstmeas}
F^{\text{st}}_{ab}(\tau) = -\frac{1}{\beta}\Bigg\langle\sum_{\alpha,\beta=1}^{k}
\sum_{j}n_{j}(\tau_{\alpha}^{e}) U_{ja}M_{\beta\alpha}\delta^{-}(\tau,\tau_{\alpha}^{e}-\tau_{\beta}^{s})\notag\\
\times\delta_{a,\lambda{\alpha}}\delta_{b,\lambda'_{\beta}}
\Bigg\rangle_{\text{MC}}.
\end{align}
The density $n_{j}(\tau_{\alpha}^{e})$ needs to be evaluated at all segment end times of segments with flavors $\lambda_{\alpha}\neq j$. For $\lambda_{\alpha}=j$ the contribution is zero since $U_{jj}=0$. Hence there is no ambiguity in how to evaluate the density exactly at the operator position.

The measurement formula for the retarded part of the improved estimator is
\begin{align}
\label{fretmeas}
F^{\text{ret}}_{ab}(\tau)\! =\! -\frac{1}{\beta}\Bigg\langle\!\sum_{\alpha,\beta=1}^{k}\!\! I(\tau_{\alpha}^{e}) M_{\beta\alpha}
\delta^{-}(\tau,\tau_{\alpha}^{e}-\tau_{\beta}^{s})
\delta_{a,\lambda{\alpha}}\delta_{b,\lambda'_{\beta}}\!\!
\Bigg\rangle_{\!\!\text{MC}},
\end{align}
where the interaction-density integral $I(\tau_{\alpha}^{e})$ can be evaluated explicitly for any configuration as follows:
\begin{align}
\label{uintegral}
I(\tau_{\alpha}^{e})&\Let\sum_{j}\int_{0}^{\beta}d\tilde{\tau}n_{j}(\tilde{\tau})U_{\text{ret}}(\tilde{\tau}-\tau_{\alpha}^{e}) \notag\\
&= -2K'(0^{+}) -\sum_{j}\sum_{\beta_{j}} s_{\beta_{j}}K'(\tau_{\beta_{j}}-\tau_{\alpha}^{e}).
\end{align}
Here it has again been used that the density changes its value only at the kink positions and is piecewise constant. The second sum in the second line is over all operators with flavor $j$ and the sign $s_{\beta_{j}}$ is positive (negative) for a creator (annihilator) as before. Since the retarded interaction also couples two segments with the same flavor, the sum in the second line contains one potentially ambiguous term: for $\tau_{\beta_{j}}=\tau_{\alpha}^{e}$ the argument of $K'$ vanishes and $K'$ is discontinuous at zero. The ambiguity is resolved by noting that the integral is over segments and therefore the time difference $\tilde{\tau}-\tau_{\alpha}^{e}$ is always negative. Hence the term $K'(0)$ means $K'(0^{-})$.
Since it is more convenient for implementation purposes, the above formula has been written such that the $K(0)$ term is to be interpreted as $K'(0^{+})$. Consequently, the difference $K'(0^{-})-K'(0^{+} )=-2K'(0^{+})$ has been separated explicitly in this formula. The term stems from the fact that $n_{j}$ does not commute with the operator $c(\tau_{\alpha}^{e})$ on the same orbital.
It corresponds to the static ($\iom=0$) component of the retarded part of the interaction $U_{\text{ret}}(\iom)$. According to Eq. \eqref{Usolver}, it may be taken into account by replacing the bare value $U$ in $F^{\text{st}}$, Eq. \eqref{fst}, by its screened value $U_{\text{scr}}$.

The integral $I(\tau_{\alpha}^{e})$ \eqref{uintegral} can be computed efficiently for a given configuration from the derivative $K'$ of the retarded interaction kernel $K$. Hence the function $K'$ as well as $K$ are passed as input to the solver\footnote{While $K'$ can be computed from the knowledge of $K$, it should be avoided to compute it inside the solver for accuracy reasons, since $K$ is usually represented on a discrete grid.}. The same integral appears in the improved estimators for the vertex functions, as discussed below. It is therefore convenient to precompute it for all times $\tau_{\alpha}^{e}$ of a configuration for which a measurement is to be performed and reuse it in the different measurements.

Note that the Green's function and the improved estimators can be measured directly in any basis, such as in imaginary time, on Matsubara frequencies or in terms of Legendre polynomials, by appropriately transforming the measurement rules. For example, in order to measure the correlation function on Matsubara frequencies, the measurement rules can be Fourier transformed. This simply amounts to replacing $\tau$ by $\inu$ and $\delta^{-}(\tau,\tau_{\alpha}^{e}-\tau_{\beta}^{s})$ by $\exp[\inu(\tau_{\alpha}^{e}-\tau_{\beta}^{s})]$ in Eqs. \eqref{gmeas}, \eqref{fstmeas} and \eqref{fretmeas}.

\begin{figure}[t]
\begin{center}
\includegraphics[scale=0.7,angle=0]{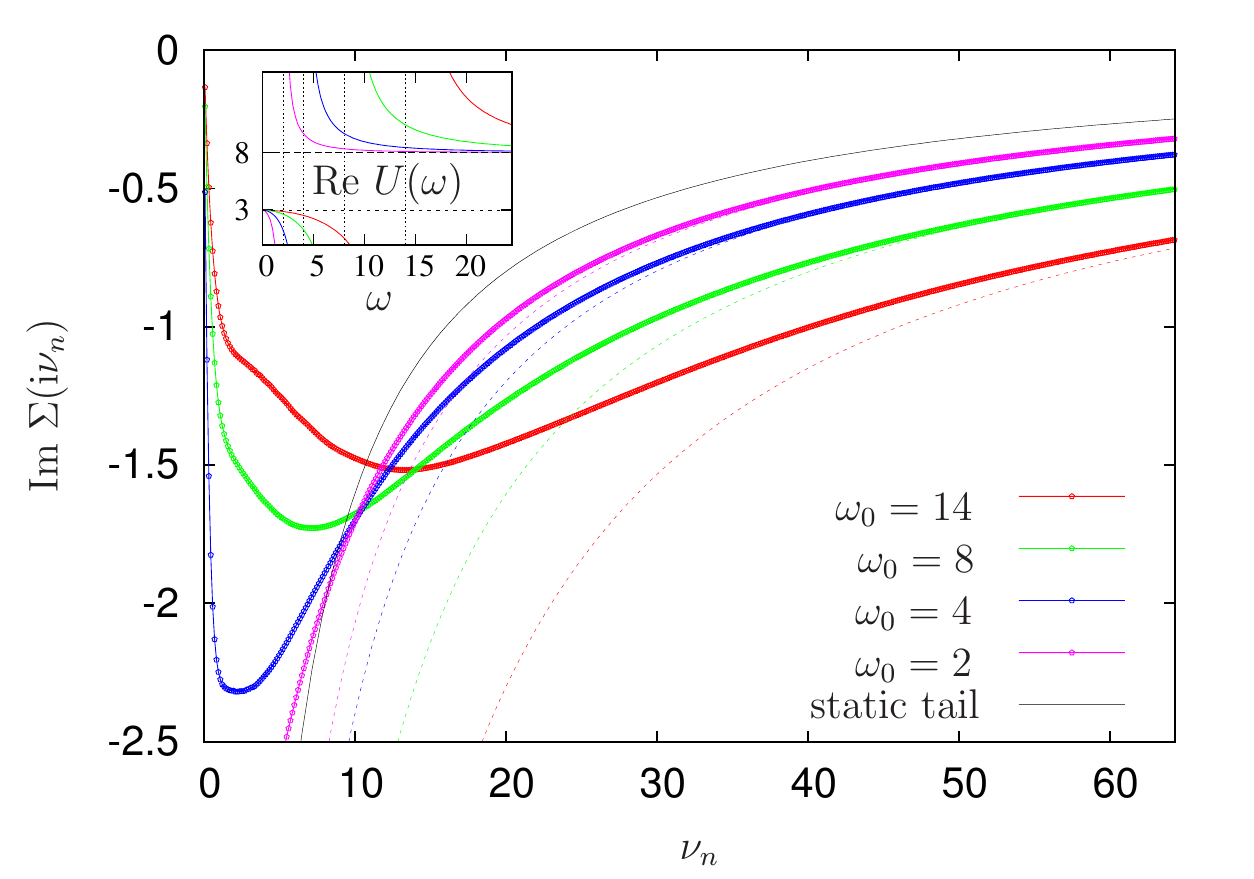} 
\end{center}
\caption{\label{fig:sigmaplasmon} (Color online) Self-energies measured using the improved estimator in the Legendre basis for different screening frequencies $\omega_{0}$ at fixed screened interaction $U_{\text{scr}}=3$. The static interaction was chosen to be $U=8$ and the temperature $T=0.02$. Thin lines show the high-frequency tails obtained from Eq. \eqref{sigma_tail1}.
The self-energy tail for a purely static interaction $U^{2}\av{\hat{n}_{\downarrow}}(1-\av{\hat{n}_{\downarrow}})$ with $U=8$ is shown for comparison. The inset shows the real part of the retarded interaction on real frequencies. Energies are given in units of the hopping $t$.
}
\end{figure}

\begin{figure}[t]
\begin{center}
\includegraphics[scale=0.7,angle=0]{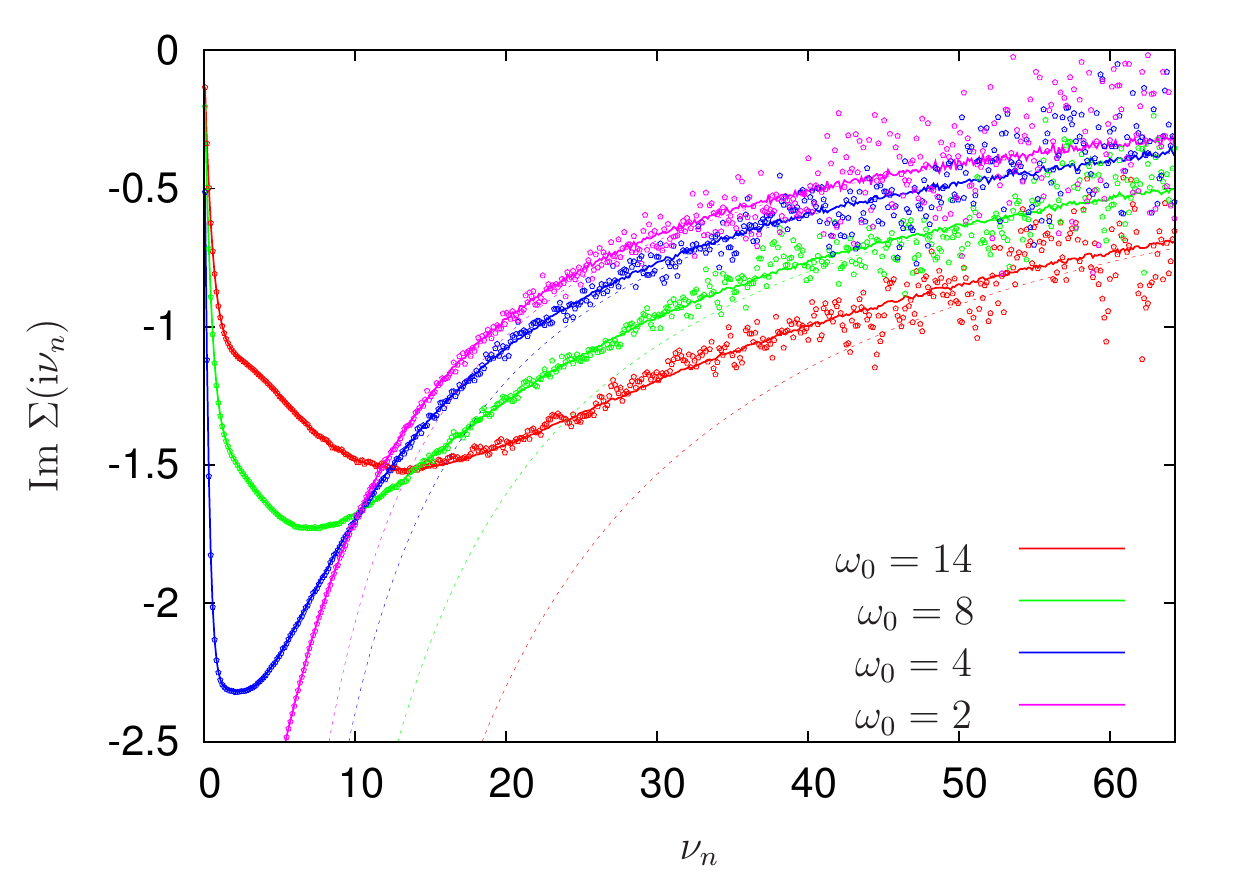} 
\end{center}
\caption{\label{fig:sigma_plasmon_dyson} (Color online) Comparison of self-energies obtained from the improved estimator (solid lines) and Dyson's equation (points) for the same model and parameters as in Fig. \ref{fig:sigmaplasmon}. The correlation functions have been measured in the same simulation and on Matsubara frequencies in both cases. Comparison with the high-frequency tails (dotted lines) reveals that the use of Dyson's equation introduces large errors in a region where the self-energy clearly has not reached its asymptotic behavior.}
\end{figure}

For low frequencies, the improved estimator can be tested by comparing the results to the ones obtained from Dyson's equation. For high frequencies, the correctness is difficult to judge due to the noise in the latter result. The high-frequency behavior of the self-energy can however be computed explicitly.
As shown in Appendix~\ref{app:tails}, the asymptotic behavior of the self-energy in the presence of the retarded interaction $U_{ij}(\tau) \Let U_{ij}\delta(\tau) +U_{\text{ret}}(\tau)$
is governed by
\begin{align}
\label{sigma_tail}
\Sigma_{a}(\inu) = \Sigma_{a}^{0} + \frac{\Sigma_{a}^{1}}{\inu} + \mathcal{O}[\frac{1}{(\inu)^{2}}],
\end{align}
where
\begin{align}
\label{sigma_tail0}
\Sigma_{a}^{0} =&  \sum_{j} \int_{0}^{\beta} d\tau U_{aj}(\tau) \av{n_{j}},
\end{align}
\begin{align}
\label{sigma_tail1}
\Sigma_{a}^{1} =& -U_{\text{ret}}(0^{+}) 
+\sum_{ij} \int_{0}^{\beta}d\tau \int_{0}^{\beta}d\tau' U_{ai}(\tau)U_{aj}(\tau')\notag\\
&\qquad\qquad\qquad\times \Big(\av{n_{i}(\tau)n_{j}(\tau')}-\av{n_{i}}\av{n_{j}}\Big).\end{align}
From the Hartree shift and $U(\iom=0)=U_{\text{scr}}$ one finds the condition for half-filling, i.e. $\mu_{1/2}=\frac{1}{2}\sum_{j}U^{\text{scr}}_{aj}$. The first term in $\Sigma^{1}$ stems from the bosonic bath and is typically the dominant contribution. Note that for a single-band model with static interaction $U$, Eq. \eqref{sigma_tail} reduces to the familiar result $\Sigma^{1}_{\uparrow}  =  U^{2}\av{n_{\downarrow}}(1-\av{n_{\downarrow}})$.

First benchmark results for the imaginary part of the self-energy obtained from the improved estimator measured in the Legendre basis~\cite{Boehnke12} are shown in Fig. \ref{fig:sigmaplasmon}. The Legendre filter eliminates the residual Monte Carlo noise and the results are seen to accurately reproduce the asymptotic behavior of the self-energy at high frequencies. The tail computed from the expression $U^{2}\av{\hat{n}_{\downarrow}}(1-\av{\hat{n}_{\downarrow}})$ is plotted for comparison for $U=8$. As expected from Eq. \eqref{sigma_tail}, one can see that the tail in the presence of the screened interaction is considerably enhanced compared to the one for the bare interaction and that the enhancement increases with increasing screening frequency $\omega_{0}$. The regime where the self-energy approaches its asymptotic behavior moves to larger frequencies with increasing $\omega_{0}$.
For the largest $\omega_{0}$, the self-energy is clearly metallic and exhibits a hump structure at small Matsubara frequencies. With decreasing screening frequency, it is strongly renormalized at small frequencies and finally exhibits insulating behavior for the smallest screening frequency. We will discuss these features in more detail in Sec.~\ref{sec:resultssigma}.

The inset of Fig.~\ref{fig:sigmaplasmon} shows the real part of the frequency dependent interaction for this model on real frequencies. For small energies, the interaction approaches its screened value $U_{\text{scr}}=3$, while for high frequencies, the interaction converges to the unscreened value $U=8$. The two regimes are separated by a pole at the respective screening frequency $\omega_{0}$.

Figure~\ref{fig:sigma_plasmon_dyson} compares the result from the improved estimator to the one obtained from Dyson's equation. The improved estimator and Green's function have been measured directly on Matsubara frequencies and within the same simulation. The large noise in the results obtained from Dyson's equation is apparent. One can also see that the error of those results is significant in a regime where the self-energy has not yet reached its asymptotic behavior. Hence the error cannot be eliminated by replacing the self-energy by its tail at high frequencies.
While these results illustrate that one can extract more accurate results for the self-energy by sampling over the same Monte Carlo configurations using the improved estimator, in practice one is interested in highest accuracy for given runtime. Therefore one needs to take into account that the measurement of the improved estimator and in particular the evaluation of the interaction-density integral, Eq. \eqref{uintegral}, slow down the simulation (depending on the perturbation order). However in practice the qualitative picture remains very similar as in Fig.~\ref{fig:sigma_plasmon_dyson} and hence the use of the improved estimator outweighs the slowdown. When measuring vertex functions, the additional overhead is completely negligible.

While the improved estimator is generally more accurate at intermediate to high energies, it should be noted that deviations may occur at low frequencies in the insulating phase because the overlap between segments and hence the statistics for $F$ are suppressed. The statistical errors of $F$ (and $G$) should therefore be monitored.

\subsection{Bosonic self-energy}
\label{sec:polarization}

Another quantity of interest is the bosonic self-energy of the impurity. Consider an action which is obtained from the impurity action \eqref{action1}  by substituting the retarded part \eqref{sret} by an action of the form
\begin{align}
S_{\text{Boson}}
=&-\frac{1}{2}\int_{0}^{\beta}d\tau\int_{0}^{\beta}d\tau' \tilde{\phi}(\tau)\mathcal{D}^{-1}(\tau-\tau')\tilde{\phi}(\tau')\notag\\
&+ \sum_{i}\int_{0}^{\beta}d\tau \tilde{\phi}(\tau)\bar{n}_{i}(\tau),
\end{align}
with the bare bosonic propagator $\mathcal{D}$ and $\bar{n_{i}}=n_{i}-\av{n_{i}}$.
This viewpoint is useful e.g. when studying the Anderson-Holstein model. The corresponding action also appears as the EDMFT effective action~\cite{Sun02,Ayral13}. By integrating out the bosonic field, it can be brought into the form \eqref{sret}, but with $n_{i}$ replaced by $\bar{n}_{i}$ (this merely amounts to a shift in the chemical potential).
For this action, the boson propagator is defined as
\begin{align}
D(\tau-\tau')\Let-\av{\tilde{\phi}(\tau)\tilde{\phi}(\tau')}.
\end{align}
In Appendix~\ref{app:polarization} it is shown that the associated self-energy $\Pi=\mathcal{D}^{-1}-D^{-1}$ (the impurity polarization) is given by
\begin{align}
\Pi(\iom) = \frac{\chi(\iom)\mathcal{D}(\iom)}{D(\iom)},
\end{align}
with the susceptibility defined in \eqref{chi}. Hence the bosonic self-energy is computed as a ratio of observables. 

\subsection{Improved estimator for the two-particle vertex}
\label{sec:impestgamma}

The impurity model two-particle vertex function finds its application in the computation of susceptibilities within DMFT~\cite{Georges96}. Diagrammatic extensions of DMFT~\cite{Toschi07,Rubtsov08,Rubtsov12,Rohringer13} also rely on the computation of a suitable (reducible or irreducible) impurity vertex function.
In this section, we discuss the computation of the reducible two-particle impurity vertex using improved estimators.

In order to unify the notation, the Green's functions (and vertices) are labeled by an  index ``$(n)$'', where $n$ denotes the number of legs. For example, the two-particle (four-leg) Green's function will hence be denoted $G^{(4)}$ and the associated improved estimator $F^{(4)}$.

The four-leg vertex function is defined by
\begin{align}
\gamma^{(4)}_{abcd}(\inu,\inu',\iom)\! =\!\!\!& \sum_{a'b'c'd'}
G^{-1}_{aa'}(\inu)G^{-1}_{cc'}(\inu'+\iom)\times\notag\\
&G^{(4),\text{con}}_{a'b'c'd'}(\inu,\inu',\iom) G^{-1}_{b'b}(\inu+\iom)G^{-1}_{d'd}(\inu'),
\end{align}
where the connected part of the two-particle Green's function $G^{(4)}_{abcd}(\inu,\inu',\iom)$ is
\begin{align}
G^{(4),\text{con}}_{abcd}(\inu,\inu',\iom) =& G^{(4)}_{abcd}(\inu,\inu',\iom) \notag\\
&- \beta[G_{ab}(\inu)G_{cd}(\inu')\delta_{\omega,0}\notag\\
&\qquad - G_{ad}(\inu)G_{cb}(\inu+\iom)\delta_{\nu\nu'}].
\end{align}
The two-particle Green's function is given by the impurity average
\begin{align}
\label{chi4def}
G^{(4)}_{abcd}(\inu,\inu',\iom) & \Let \av{c_{a}(\inu)c^{*}_{b}(\inu+\iom)c_{c}(\inu'+\iom)c^{*}_{d}(\inu')}.
\end{align}
Its relation to the four-leg vertex function is depicted diagrammatically in Fig.~\ref{vertices}. Similarly as for the self-energy, it is more reliable to compute the vertex from improved estimators instead of using the above definition. Using the equation of motion, one can derive an expression for the connected part  of the two-particle Green's function:
\begin{align}
\label{chi4con}
G^{(4),\text{con}}_{abcd}(\inu,\inu',\iom)=&
\sum_{i} G_{ai}(\inu) F^{(4)}_{ibcd}(\inu,\inu',\iom)
\notag\\
&-\sum_{i} F'_{ai}(\inu)\, G^{(4)}_{ibcd}(\inu,\inu',\iom),
\end{align}
where  $F'=G\Sigma$ as before\footnote{For a diagonal basis, $F'=F$. In general, $F'$ can be either computed from $F$ and $G$, or measured directly. In the latter case, the density and the interaction-density integral \eqref{uintegral} have to be evaluated at the creator times $\tau_{\beta}^{s}$.}.
$F^{(4)}_{abcd}(\inu,\inu',\iom)$ consists of a static and a retarded part:
\begin{align}
\label{hstatic}
&F^{(4),\text{st}}_{abcd}(\tau_{a},\tau_{b},\tau_{c},\tau_{d}) =\notag\\
&\sum_{j} \av{n_{j}(\tau_{a})U_{ja}c_{a}(\tau_{a})c_{b}^{*}(\tau_{b})c_{c}(\tau_{c})c_{d}^{*}(\tau_{d})},\\
\label{hretfinal}
&F^{(4),\text{ret}}_{abcd}(\tau_{a},\tau_{b},\tau_{c},\tau_{d}) =\notag\\
&\int_{0}^{\beta}d\tilde{\tau}\sum_{i}\av{n_{i}(\tilde{\tau})U_{\text{ret}}(\tilde{\tau}-\tau_{a}) c_{a}(\tau_{a})c_{b}^{*}(\tau_{b})c_{c}(\tau_{c})c_{d}^{*}(\tau_{d})}.
\end{align}
The improved estimators for the vertex function are measured similarly to Eqs. \eqref{fstmeas} and \eqref{fretmeas}. The main difference is that the matrix $M$ in these formulas is replaced by an antisymmetrized product of two $M$-matrices.
The measurement formula for the static contribution to the improved estimator for the four-leg vertex function was given in Ref.~\onlinecite{Hafermann12}. The contribution from the retarded interaction is obtained by replacing the prefactor $\sum_{j}n_{j}(\tau_{\alpha}^{e})U_{ja}$ by the interaction density integral $I(\tau_{\alpha}^{e})$. 
On Matsubara frequencies, $G^{(4)}_{abcd}(\inu,\inu',\iom)$ is measured in two steps. In a first step, the Fourier transform
\begin{align}
\label{mfourier}
M_{ab}(\inu,\inu')= \sum_{\alpha\beta=1}^{k} M_{\beta\alpha}e^{\inu\tau_{\alpha}^{e}}e^{-\inu'\tau_{\beta}^{s}}\delta_{a\lambda_{\alpha}}\delta_{b\lambda'_{\beta}}
\end{align}
is computed. The final measurement is constructed as the antisymmetrized product thereof and similarly for the correlators \eqref{hstatic} and \eqref{hretfinal}.
With $N_{\nu}$ being the number of fermionic/bosonic frequencies, this measurement scales as $k^{2} N_{\nu}^{2} + N_{\nu}^{3}$ for the two steps instead of $k^{4} N_{\nu}^{3}$.
The improvement of the results by using the improved estimator for the vertex is comparable to that reported in Ref.~\onlinecite{Hafermann12}.

\subsection{Improved estimator for the electron-boson vertex}
\label{sec:impestlambda}

The three-leg vertex is of interest because it contains information on the effective electron boson interaction\footnote{Integrating out the bosonic field from the correlation function \av{cc^{*}\phi} using \eqref{intphi}, leads to the correlation function \av{cc^{*}n}, from which the vertex is derived.}
and it is relevant for the recently proposed dual boson method~\cite{Rubtsov12}. Here we discuss the computation of the three-leg vertex function as defined in the dual boson approach.

\begin{figure}[t]
\begin{center}
\includegraphics[scale=0.6,angle=0]{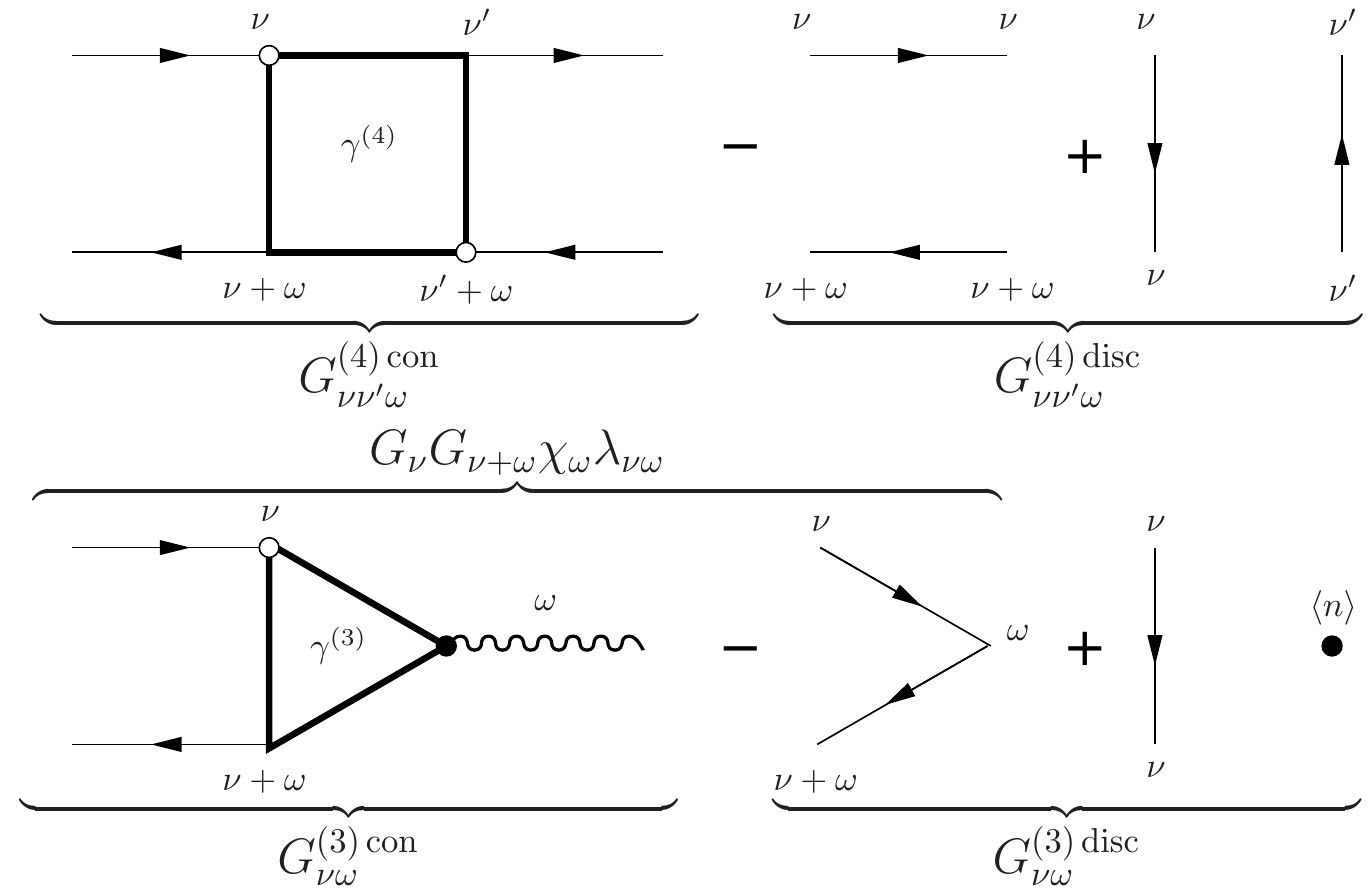} 
\end{center}
\caption{\label{vertices} Diagrammatic representation of the four-leg (three frequency) correlation function $G^{(4)}$ [Eq. \eqref{chi4def}] (upper panel) and three-leg (two frequency) correlation function $G^{(3)}$ [Eq. \eqref{chi3def}] (lower panel) in terms of their connected and disconnected parts. Spin indices are omitted. Straight lines with arrows denote fully dressed propagators. The wavy line represents the charge susceptibility. The juxtaposition illustrates the formal difference between the vertex functions $\gamma^{(4)}$ and $\lambda$.
}
\end{figure}

\begin{figure}[b]
\begin{center}
\includegraphics[scale=0.7,angle=0]{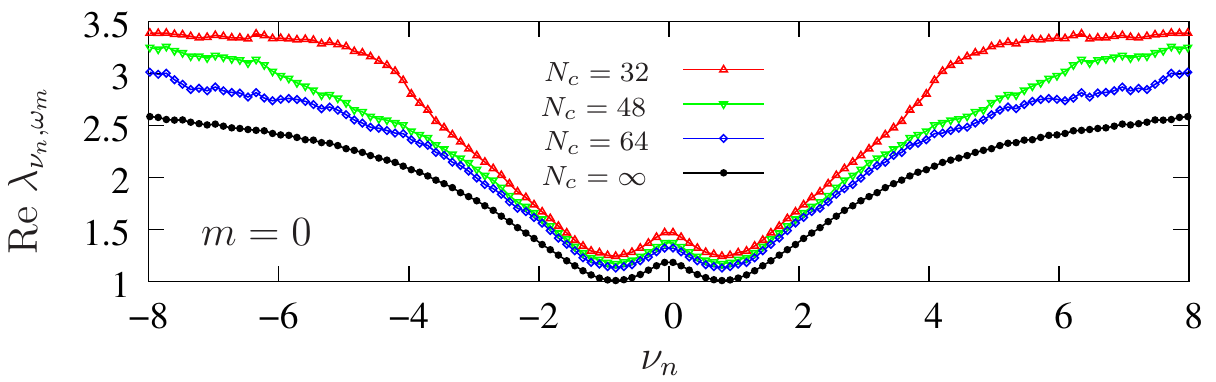} 
\end{center}
\caption{\label{fig:lambda_plasmon_cutoff} (Color online) Three-leg vertex for fixed bosonic frequency $\omega_{m}=0$ computed from the four-leg vertex using the identity \eqref{lambdafromgamma} for a given cutoff $N_{c}$ of the frequency sum. The result is compared to the one obtained from the improved estimator \eqref{impestlambda} corresponding to an infinite cutoff. The convergence with the cutoff is slow, leading to a large systematic error for all frequencies.
}
\end{figure}

The three-leg correlation function is defined as
\begin{align}
\label{chi3def}
G^{(3)}_{abc}(\inu,\iom)&\Let -\av{c_{a}(\inu)c^{*}_{b}(\inu+\iom)n_{c}(\iom)}.
\end{align}
In analogy to the four-leg vertex, its connected part is given by
\begin{align}
G_{abc}^{(3),\text{con}}(\inu,\iom)\Let G^{(3)}_{abc}(\inu,\iom) -& [\beta G_{ab}(\inu+\iom)\av{n_{c}}\delta_{\omega,0}\notag\\
&-G_{ac}(\inu+\iom)G_{cb}(\inu)].
\end{align}
In terms of the connected part of the three-leg correlation function, the vertex $\lambda$ in the charge channel as defined in Ref.~\onlinecite{Rubtsov12} is computed as
\begin{align}
\label{lambdafromcon}
\lambda_{ab}(\inu,\iom)=\frac{1}{\chi(\iom)}\Big(&\sum_{a'b'c'}G_{aa'}^{-1}(\inu)G_{b'b}^{-1}(\inu+\iom)\notag\\
&\qquad\times G^{(3),\text{con}}_{a'b'c'}(\inu,\iom)-1\Big).
\end{align}
The relation between these quantities is depicted diagrammatically in the lower panel of Fig. \ref{vertices}. Although they are not defined completely analogously, the three-leg and four-leg vertices are closely related. For a single-band model, the identity between the two vertices reads~\cite{Rubtsov12}
\begin{align}
\label{lambdafromgamma}
\lambda_{\sigma}(\inu,\iom) = \frac{1}{\chi(\iom)} &\Big(\frac{1}{\beta}\sum_{\nu'\sigma'}\gamma^{(4)}_{\sigma\sigma\sigma'\sigma'}(\inu,\inu',\iom)\notag\\
&\quad\times G_{\sigma'}(\inu')G_{\sigma'}(\inu'+\iom)-1\Big).
\end{align}
This relation is however problematic to use numerically, because the frequency sum converges slowly, as illustrated in Fig. \ref{fig:lambda_plasmon_cutoff}. The three-leg vertex is plotted for different cutoffs of the frequency sum in \eqref{lambdafromgamma} and a direct determination from the three-leg correlation function which corresponds to an infinite cutoff. One can see that above the cutoff frequency, the tail is unreliable. Even for a relatively high cutoff, a large systematic error remains. In addition, the results exhibit more noise than the determination of $\lambda$ through its improved estimator, although an improved estimator was used to obtain the four-leg vertex. The three-leg vertex should therefore be computed directly from the corresponding three-leg correlation function, as discussed in the following.

The derivation of the improved estimator for the three-leg vertex is sketched in Appendix~\ref{app:impest}. One obtains the following relation for the connected part of the three-leg correlation function in complete analogy to Eq. \eqref{chi4con}:
\begin{align}
\label{impestlambda}
G^{(3),\text{con}}_{abc} (\inu,\iom) 
=& \sum_{i} G_{ai}(i\nu)F^{(3)}_{ibc}(\inu,\iom)
\notag\\
&-\sum_{i}F'_{ai}(\inu)G^{(3)}_{ibc} (\inu,\iom),
\end{align}
where $F'$ is defined as before. The static and retarded contributions to the improved estimator $F^{(3)}$ read
\begin{align}
&F^{(3),\text{st}}_{abc}(\tau_{1},\tau_{b},\tau_{c}) \Let\notag\\
& -\sum_{j}\av{T_{\tau} n_{j}(\tau_{a})U_{ja}c_{a}(\tau_{a})c_{b}^{*}(\tau_{b})n_{c}(\tau_{c})},\notag\\
&F^{(3),\text{ret}}_{abc}(\tau_{1},\tau_{b},\tau_{c}) \Let\notag\\
&-\int_{0}^{\beta}d\tilde{\tau}\sum_{i}\av{T_{\tau} n_{i}(\tilde{\tau})U_{\text{ret}}(\tilde{\tau}-\tau_{a}) c_{a}(\tau_{a})c_{b}^{*}(\tau_{b})n_{c}(\tau_{c})}.
\end{align}

As in the four-leg case, the above correlation functions are measured by multiplying the measurement of the corresponding Green's function [\eqref{chi3def} in this case] by  $\sum_{j}n_{j}(\tau_{\alpha}^{e})U_{ja}$ and the interaction-density integral $I(\tau_{\alpha}^{e})$, respectively. The frequency measurement of the three-leg Green's function can in turn be written as the product of the Fourier transform of the $M$-matrix, Eq. \eqref{mfourier}, and the Fourier transform of the density, Eqs. \eqref{nw1} and \eqref{nw2}. The former is also required to measure the four-leg correlation functions and the latter for the measurement of the density-density correlation function. When these quantities are measured, the three-leg function $G^{(3)}$ can therefore be obtained at negligible additional computational cost.

\section{Results}
\label{sec:results}

In the following, we present results obtained using the improved measurements. The test case has been defined in Sec.~\ref{sec:impesttest}.

\subsection{Self-energy}
\label{sec:resultssigma}

\begin{figure}[t]
\begin{center}
\includegraphics[scale=0.7,angle=0]{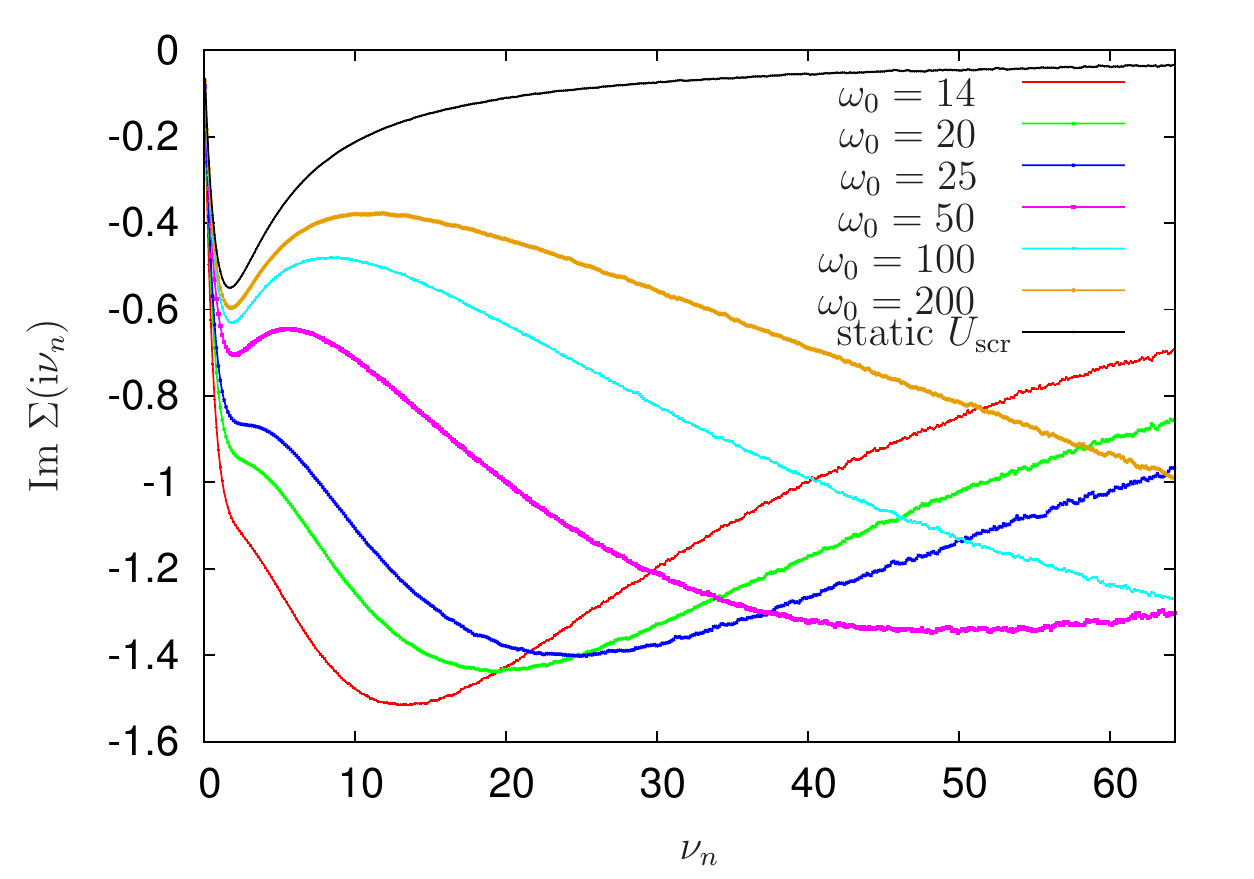} 
\end{center}
\caption{\label{fig:sigma_plasmon_antiadiabatic} (Color online) Self-energy as a function of Matsubara frequencies in the antiadiabatic regime. The parameters are otherwise the same as in Fig. \ref{fig:sigmaplasmon}.
Note that the asymptotic regime, where the self-energy is described by Eqs.~\eqref{sigma_tail}-\eqref{sigma_tail1} lies outside the plot range for the largest  screening frequencies.
 }
\end{figure}

In order to better understand the behavior of the self-energy in the presence of a retarded interaction, it is useful to consider the opposite regimes of large and small screening frequency in more detail.
The so-called antiadiabatic regime is characterized by $\omega_{0} > U-U_{\text{scr}}$. This situation appears to be common to plasmonic screening in real materials (see, e.g., Ref.~\onlinecite{Casula12} and references therein). Results for the self-energy in this regime are shown in Fig.~\ref{fig:sigma_plasmon_antiadiabatic}. For energies small compared to the screening frequency $\omega_{0}$, the electrons essentially experience the screened interaction.
In order to see this, the result from a calculation with a purely static interaction equal to $U_{\text{scr}}$ is plotted in this figure for comparison.
This is precisely the situation one commonly considers in calculations for real materials which neglect the effect of a retarded interaction: The Hubbard interaction $U$ is taken equal to the anticipated screened value. However, one can see that the static model delivers a rather poor description even of the low-energy behavior of the self-energy unless the screening frequency is extremely high. The static approximation only becomes exact in the limit of infinite screening frequency.

Following the evolution of the self-energies with decreasing $\omega_{0}$, one can see that for all screening frequencies considered in this figure, a feature reminiscent of the minimum in the self-energy of the static calculation remains, which degenerates to a hump structure for the smallest frequency considered. This feature is likely to carry information on the scale of the screened interaction $U_{\text{scr}}$.
One also observes a steep increase of the self-energy at small energies for smaller screening frequencies as well as a minimum at the scale of the screening frequency $\omega_{0}$. These features translate to corresponding features previously reported for the spectral functions for models including a retarded interaction~\cite{Werner10,Casula12,Casula12-2}: a suppression of the spectral weight at low energy and the appearance of a plasmon satellite at (and at multiples of) the screening frequency. Note that some of the data in Figs.~\ref{fig:sigma_plasmon_antiadiabatic} and \ref{fig:sigma_plasmon_adiabatic} corresponds and may be directly compared to the spectral functions reported in the bottom panel of Fig. 3 of Ref.~\onlinecite{Werner10}.

One finally observes that all self-energies in this figure extrapolate to very similar values at zero, showing that the scattering rate is essentially independent of the frequency dependence of the interaction and rather determined by its low-frequency screened value. The density of states at the Fermi level hence remains essentially unchanged with respect to the static calculation, in agreement with previous results.~\cite{Casula12} This is confirmed by inspection of the imaginary-time Green's functions at $\tau=\beta/2$, which almost coincide\footnote{In the low temperature limit, $(\beta/2)G(\beta/2)$ may be used as an approximate measure of the density of states at the Fermi level.} (not shown).

\begin{figure}[t]
\begin{center}
\includegraphics[scale=0.7,angle=0]{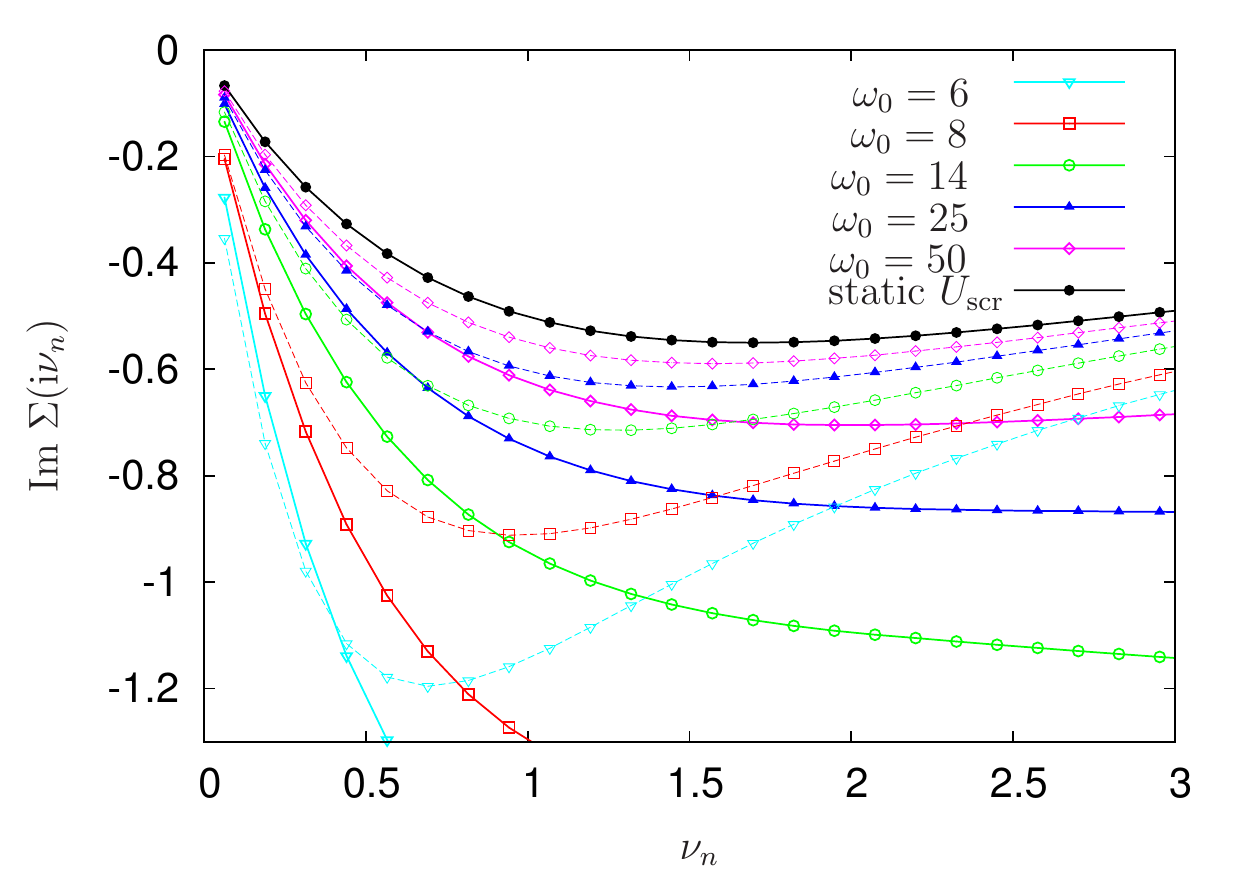} 
\end{center}
\caption{\label{fig:sigma_eff} (Color online) Comparison of the self-energy for the model with dynamical interaction (solid lines) with the result for the low-energy effective model with static interaction (dashed lines). The curve labeled ``static $U_{\text{scr}}$'' is the result for a static calculation without band renormalization, which corresponds to the exact result for $\omega_{0}\to\infty$.
The effective model description breaks down for $\omega_{0}\lesssim 8$.}
\end{figure}

As shown in Ref.~\onlinecite{Casula12-2}, for large screening frequencies, the low-energy physics of the model is approximately governed by an effective model with a purely static interaction given by $U_{\text{scr}}$, but with an additional bandwidth reduction by a factor $Z_{B}=\exp(-\lambda^{2}/\omega_{0}^2)$. The spectral weight at the Fermi level of the original model compared to that of the effective model is reduced by the same factor. The physical origin of this spectral weight transfer are processes involving the emission or absorption of one or multiple plasmons.

Spectral functions of the full model and the effective model have been compared in Ref.~\onlinecite{Casula12-2}. Here we compare the two models on the level of the self-energy. Figure~\ref{fig:sigma_eff} compares results of the original model with those of the effective model. For a given $\omega_{0}$, the self-energies approach each other for small frequencies. The approximation clearly breaks down for small screening frequencies, as can be seen for the results with $\omega_{0}=6$. Comparing to the calculation with an unrenormalized band ($Z_{B}=1$, labeled ``static $U_{\text{scr}}$'' in Fig.~\ref{fig:sigma_eff}), one can see that the bandwidth renormalization is essential to approximate the low-frequency behavior. Note that the effective model description is restricted to low energy and in particular does not reproduce the plasmon peaks at high energy.

\begin{figure}[t]
\begin{center}
\includegraphics[scale=0.7,angle=0]{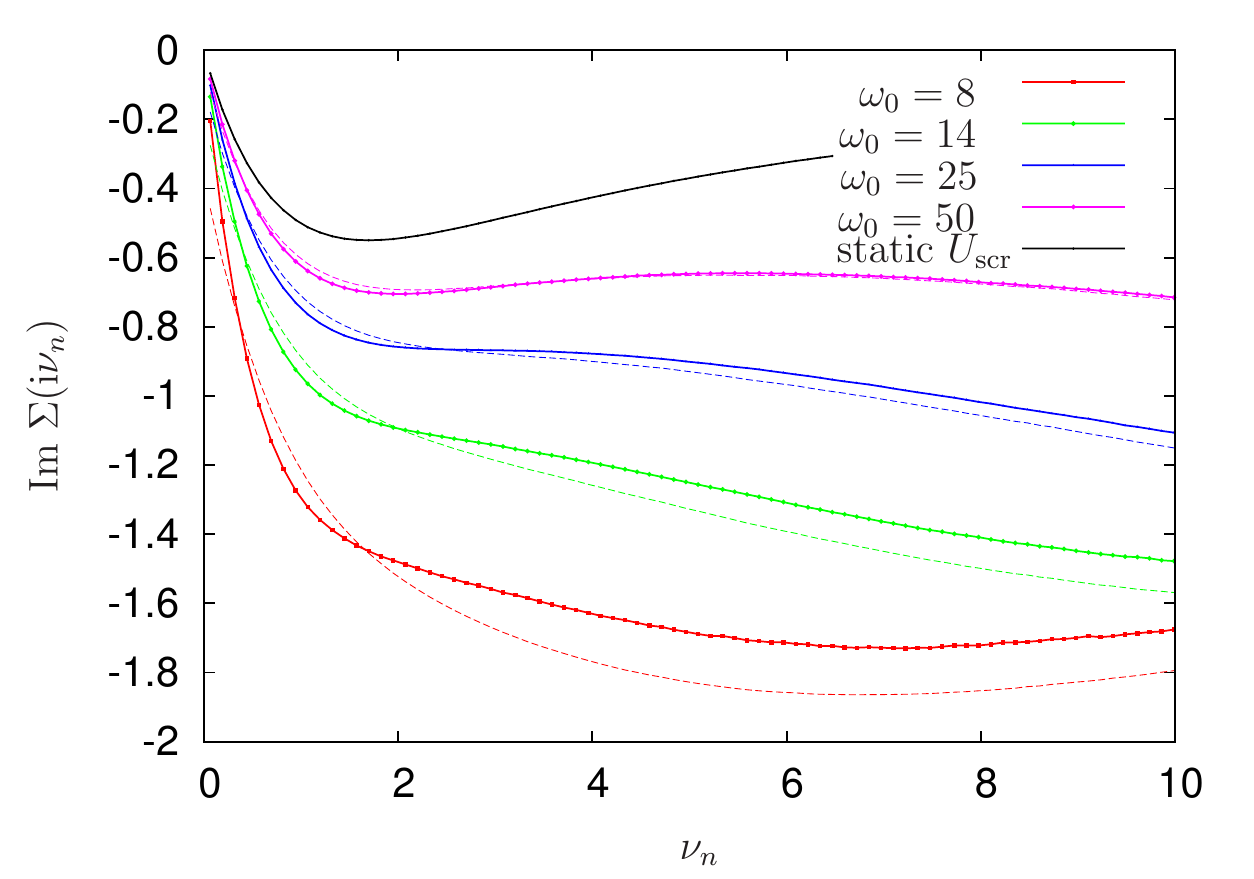} 
\end{center}
\caption{\label{fig:sigma_dala} (Color online) Comparison of the self-energy for the model with dynamical interaction (solid lines, with symbols) with the DALA result (dashed lines).}
\end{figure}

\begin{figure}[b]
\begin{center}
\includegraphics[scale=0.7,angle=0]{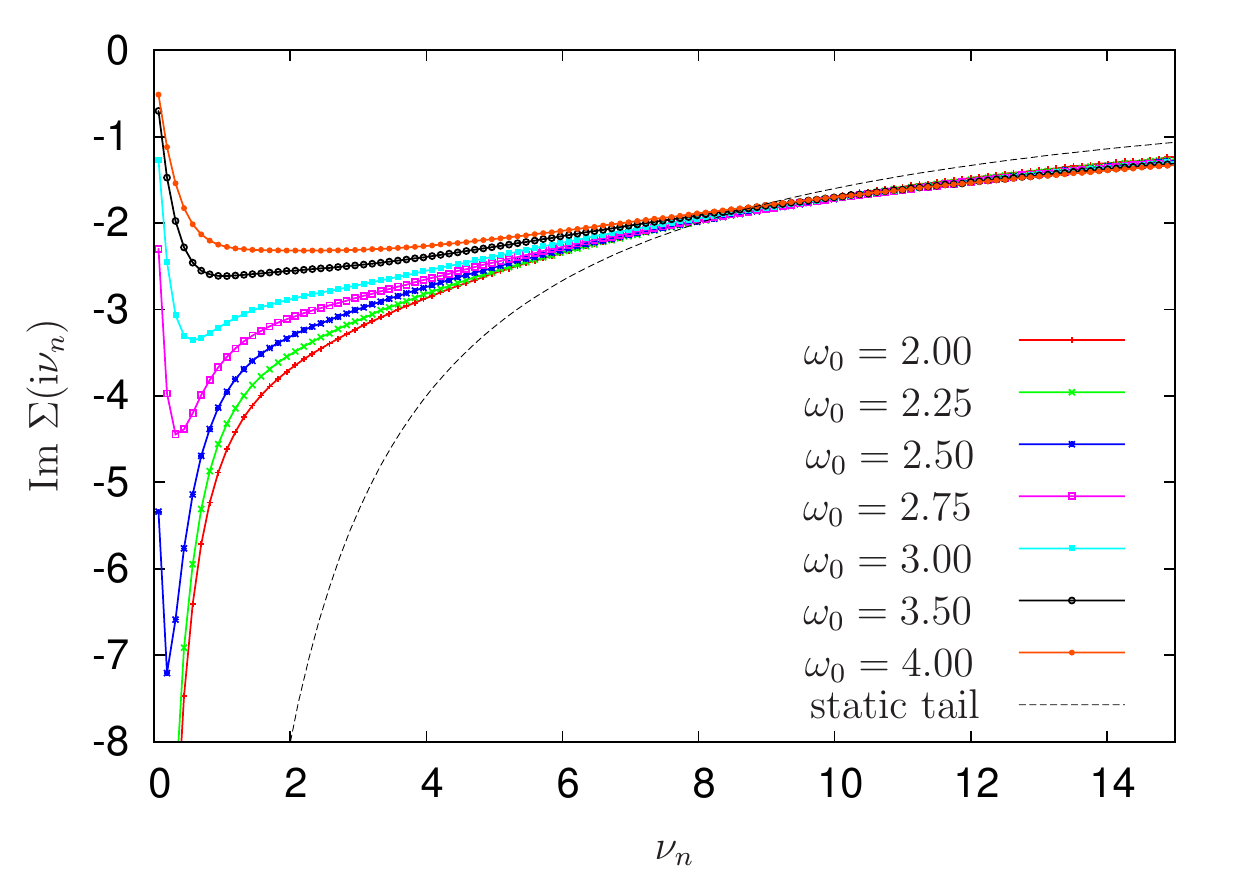} 
\end{center}
\caption{\label{fig:sigma_plasmon_adiabatic} (Color online) Self-energy as a function of Matsubara frequencies in the adiabatic regime. The static tail computed from the static unscreened interaction $U$ is shown for comparison.
}
\end{figure}

For high energies, the electrons experience a partially screened and hence much larger interaction. This leads to the strong renormalization of the self-energy for intermediate to high frequencies. This behavior and the corresponding spectral weight redistribution from low to higher energies can be described in the so-called dynamic atomic limit approximation (DALA)~\cite{Casula12}. The approximation relies on the separation of the low-energy scale set by $U_{\text{scr}}$ and the high-energy physics governed by the retarded part of the interaction. It is based on the following ansatz for Green's function: $G(\tau) = G_{\text{scr}}(\tau) B(\tau)$. Here $G_{\text{scr}}$ is the Green's function obtained from the calculation with a static interaction equal to $U_{\text{scr}}$. The bosonic propagator $B(\tau)$ is evaluated in the dynamical atomic limit, which yields $B(\tau)=\exp[-K(\tau)]$.

DALA results are compared to the numerically exact results on the level of self-energy in Fig. \ref{fig:sigma_dala}.
The DALA reproduces the high-energy features of the self-energy (in particular the minimum at the screening frequency observed in Fig.~\ref{fig:sigma_plasmon_antiadiabatic}) remarkably well. As expected, it works better the higher the screening frequency, i.e., when the assumption of the separation of energy scales is well justified. At low frequencies the approximation deviates because of the importance of the hybridization in determining the lowlow-energyenergy properties. As proposed in Ref.~\onlinecite{Casula12}, the DALA can be combined with the Lang-Firsov approach, i.e., with the previously introduced effective model, to cure this deficiency.

Figure~\ref{fig:sigma_plasmon_adiabatic} shows the self-energy in the opposite adiabatic regime. 
Here the screening is inefficient already at rather small energies (of the order of the plasmon frequency), so that the electrons experience the unscreened interaction.
As a result, the self-energy is strongly affected for small Matsubara frequencies and a signature on the energy scale of the screened interaction is no longer visible.
One can further see that the high-frequency behavior is similar to that obtained in a static calculation with the static interaction equal to the unscreened value $U$.

\begin{figure}[t]
\begin{center}
\includegraphics[scale=0.65,angle=0]{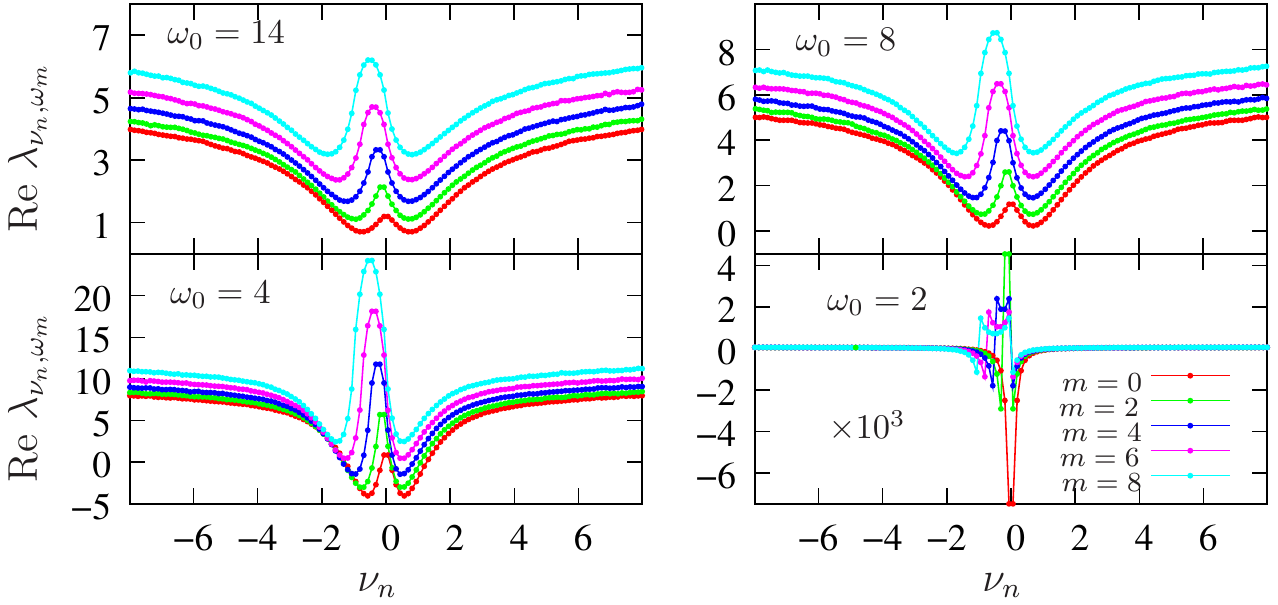} 
\end{center}
\caption{\label{fig:lambda_plasmon} (Color online) Three-leg vertex $\lambda$ for fixed bosonic frequency as a function of the fermionic frequency and for different screening frequencies $\omega_{0}$. As the Mott metal-insulator transition is approached for decreasing $\omega_{0}$, the vertex develops structure at small fermionic frequencies.
}
\end{figure}

Electrons at the Fermi level, on the other hand, experience the screened interaction $U_{\text{scr}}$. As a consequence, the self-energy displays an upturn for small frequencies and metallic behavior. If the screening frequency is sufficiently small, or the temperature is sufficiently high, the upturn is no longer resolved on the discrete Matsubara frequencies and the system behaves as an insulator. Hence, as the screening frequency decreases, a Mott metal-to-insulator transition takes place, which is first-order~\cite{Werner10}.

\subsection{Three-leg vertex}

Figure~\ref{fig:lambda_plasmon} shows results for the three-leg vertex determined from the improved estimator for various values of the screening frequency. The vertex is plotted as a function of the fermionic frequency for different bosonic frequencies.
It exhibits a peak at small frequencies, the maximum of which is shifted to higher frequencies with increasing bosonic frequency. This feature grows in magnitude as the screening frequency decreases and the Mott transition is approached. Sufficiently close to the transition, the vertex changes sign. In the Mott insulating phase the vertex is considerably larger in magnitude and the structure is different, reflecting the corresponding changes in the four-leg vertex function (see below).

\begin{figure}[t]
\begin{center}
\includegraphics[scale=0.65,angle=0]{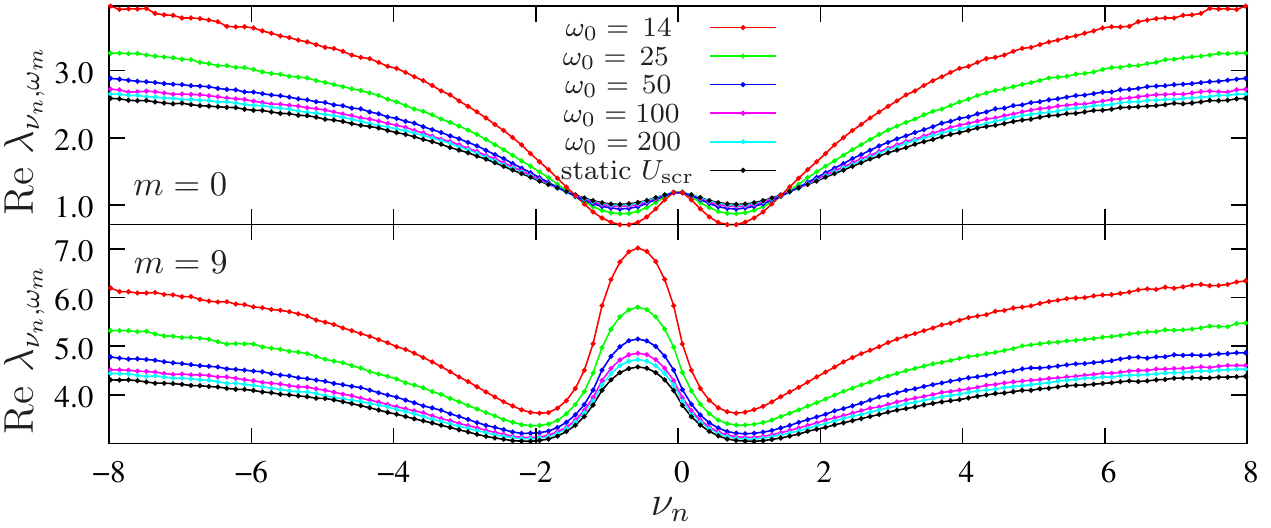} 
\end{center}
\caption{\label{fig:lambda_plasmon_antiadiab} (Color online) Three-leg vertex for two different bosonic frequencies as a function of the fermionic Matsubara frequency in the antiadiabatic regime. As for the self-energy, the result is similar to the one from a static calculation with the interaction equal to its screened value when the screening frequency is large.
}
\end{figure}

Figure~\ref{fig:lambda_plasmon_antiadiab} shows the three-leg vertex in the antiadiabatic regime. Similarly to the self-energy, the vertex is close to the result for the calculation with a static $U$ taken to be equal to the screened value and is considerably enhanced as the screening frequency decreases. The difference is larger for finite transferred frequency than for $\omega=0$. For $\omega=0$, the vertex appears to coincide at the first Matsubara frequency for all screening frequencies.

\subsection{Four-leg vertex}
\begin{figure*}[t]
\begin{center}
\includegraphics[scale=1.35,angle=0]{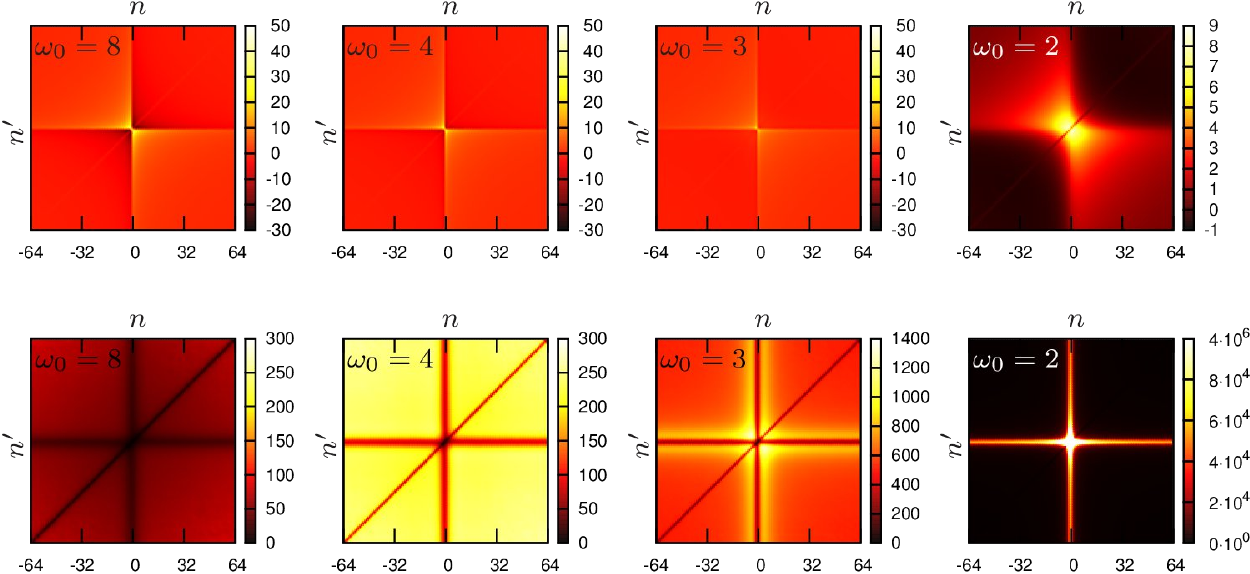} 
\end{center}
\caption{\label{fig:vertex2d} (Color online) Two-particle Green's function $\Re G_{\uparrow\uparrow}^{(4)}(\nu_{n},\nu_{n'},\omega)$ (upper panel) and vertex function $\Re \gamma_{\uparrow\uparrow}^{(4)}(\nu_{n}\nu_{n'}\omega)$ (lower panel) as a function of Matsubara frequency indices $n$ and $n'$ for fixed bosonic frequency $\omega=0$ and different screening frequencies $\omega_{0}=8,4,3,2$ (from left to right). The parameters are otherwise the same as in Fig. \ref{fig:sigmaplasmon}.
}
\end{figure*}
\begin{figure*}[t]
\begin{center}
\includegraphics[scale=1.35,angle=0]{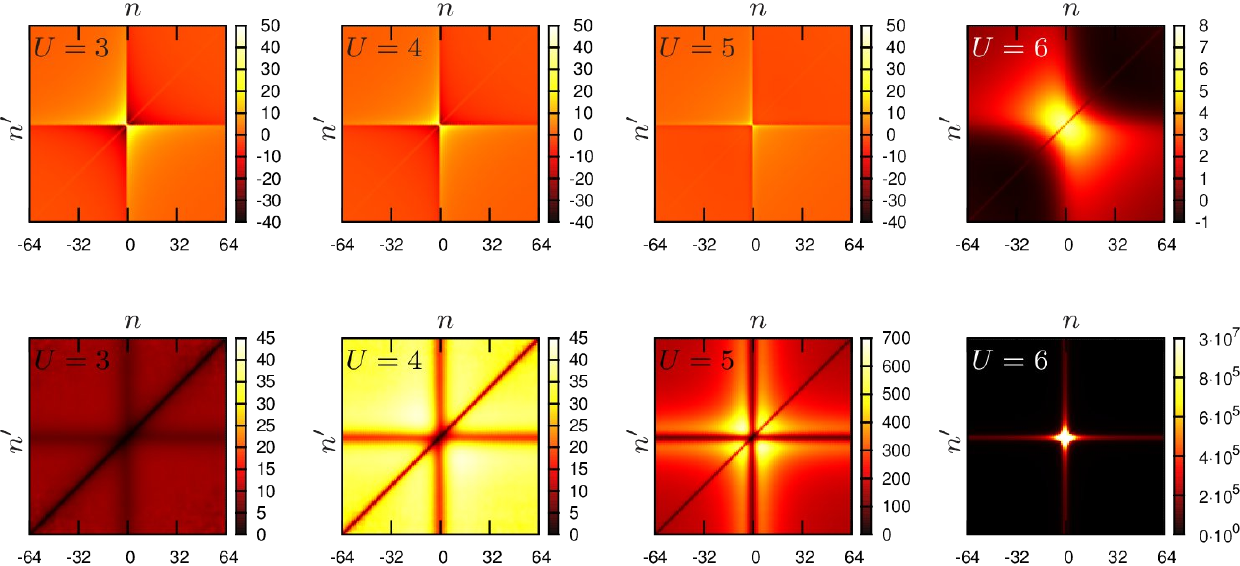} 
\end{center}
\caption{\label{fig:vertex2d_noscreening} (Color online) Two-particle Green's function $\Re G^{(4)}_{\uparrow\uparrow}(\nu_{n},\nu_{n'},\omega)$ (upper panel) and vertex function $\Re \gamma_{\uparrow\uparrow}^{(4)}(\nu_{n},\nu_{n'},\omega)$ (lower panel) plotted as in Fig. \ref{fig:vertex2d}, but computed for the model without screening for different values of the static interaction $U=3,4,5,6$ (from left to right).
}
\end{figure*}

Results for the two-particle Green's function and the vertex function obtained from the improved estimator are plotted in Fig.~\ref{fig:vertex2d} for different screening frequencies $\omega_{0}$ at half filling. Here we focus on the spin-up-up components for visualization purposes. Similar conclusions apply for the up-down components. Because of the particle-hole symmetry, both quantities are purely real. The transferred frequency is kept fixed at $\omega=0$ and  results are plotted as a function of the two fermionic frequencies. For high screening frequencies the system is metallic and the two-particle Green's function and vertex function exhibit the typical structures of the metallic phase. The structure of the two-particle Green's function is mainly determined by its disconnected part defined in Fig.~\ref{vertices}, since the vertex is comparatively small.
The cross and diagonal structures of the vertex are also present in the spin (magnetic) and  charge (density) components, where they have been observed previously~\cite{Rohringer12,Kinza13}. For small coupling these structures of the vertex can be understood in terms of perturbation theory~\cite{Rohringer12}.
As the Mott transition is approached, the contrast in the two-particle Green's function diminishes while the magnitude of the vertex function increases. Very close to the transition ($\omega_{0}=3$), the vertex functions develops peak structures at low Matsubara frequencies, in particular on the secondary diagonal $\nu'_{n}=-\nu_{n}$. In the insulator ($\omega_{0} = 2$), the vertex diverges at low frequency (for $T\to 0$), while the regions with highest intensity maintain a cross structure.
The results are in qualitative agreement with the ones of Ref.~\onlinecite{Huang13}, where an ohmic screening model was used instead of the plasmonic screening model employed here and hence do not depend on the particular form of the frequency-dependent part of the interaction.

In Fig. \ref{fig:vertex2d_noscreening}, the same quantities as in Fig. \ref{fig:vertex2d} are plotted, albeit obtained from calculations where the interaction was taken to be static. That is, $U(\iom)=U$ and $U$ was varied across the metal-insulator transition. The figure therefore shows the evolution of the two-particle Green's function and vertex across the interaction driven Mott transition in contrast to the screening frequency driven transition of Fig. \ref{fig:vertex2d}. One can see that both quantities exhibit all the qualitative features observed in the screening driven transition. In particular, the emergence of the peak structures on the secondary diagonal very close to the transition can be seen for $U/t=5$ (the Mott transition occurs at $U\approx 5.1$). The evolution of the two quantities is hence identified to be a generic feature of the Mott transition.
The similarity between the two-particle quantities for the interaction and screening driven transitions also holds for higher bosonic frequencies and the spin-up-down component as well.

\begin{figure}[t]
\begin{center}
\includegraphics[scale=0.65,angle=0]{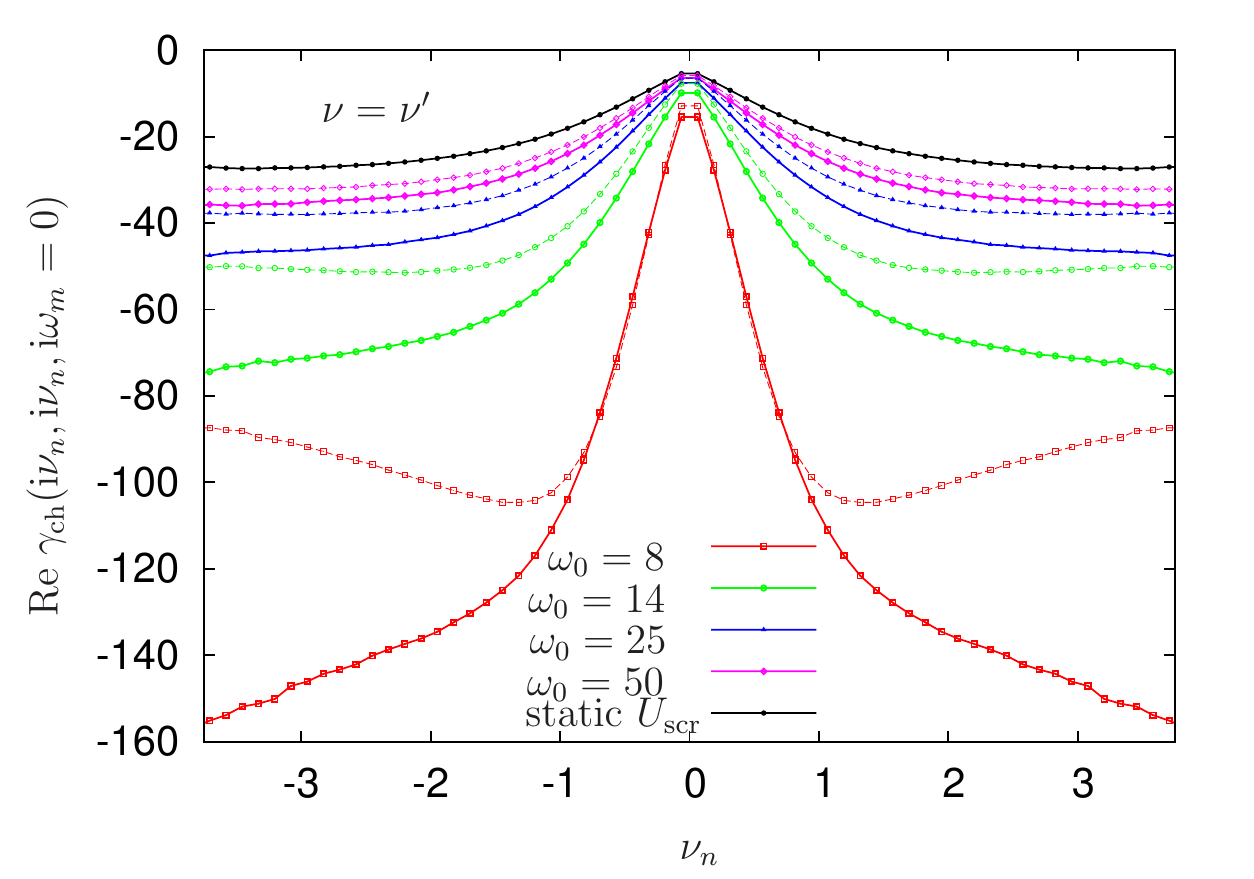} 
\end{center}
\caption{\label{fig:gammachph} (Color online) Comparison of the low-energy behavior of the vertex function in the charge channel for the model with retarded interaction (solid lines) and the low-energy effective model (dashed lines) for $\inu=\inu'$ and $\iom=0$.
}
\end{figure}

We have seen that an effective model with static interaction and a renormalized bandwidth approximately captures the low-energy behavior of the model with retarded interaction as shown for the self-energy in Fig.~\ref{fig:sigma_eff}. This can be expected to be the case also for the vertex function.
Results for the vertex should be compared in the limit where all frequencies are taken to zero. An extrapolation from the discrete Matsubara representation is difficult, but one can observe the trend already without extrapolation.  In Fig. \ref{fig:gammachph}, the vertex in the charge channel ($\gamma_{\text{ch}}\Let\gamma^{\uparrow\uparrow\uparrow\uparrow} - \gamma^{\uparrow\uparrow\downarrow\downarrow}$) is plotted for the two models for $\iom=0$ and the cut $\inu=\inu'$, which consists of contributions from the transverse particle-hole channel~\cite{Rohringer12}. Note that along this cut, $\gamma^{\uparrow\uparrow\uparrow\uparrow}_{\nu,\nu,\omega=0}=0$, which, by antisymmetry, is ultimately a consequence of the Pauli principle. Hence $-\gamma_{\text{ch}}=\gamma_{\text{sp}}\Let\gamma^{\uparrow\uparrow\uparrow\uparrow} - \gamma^{\uparrow\uparrow\downarrow\downarrow}$ along this cut.
The vertices are evidently similar at low frequencies. One can further see that they approach the limiting value for the model with a static screened interaction and unrenormalized bandwidth as the screening frequency $\omega_{0}$ is increased. At high energies, the vertices are very different, as is the case for the self-energy. The high-frequency behavior of the vertex is enhanced for the model with retarded interaction. This is expected, because the high-frequency behavior is governed by the large unscreened interaction.

\section{Conclusions and outlook}
\label{sec:conclusions}

In this paper, the technical modifications of the continuous-time hybridization expansion algorithm required to accurately compute the impurity susceptibility, self-energy and three- and four-leg vertex functions in the presence of a retarded interaction have been discussed.
The improved measurements lead to substantially more accurate results for a given runtime. This is useful for solving impurity models with retarded interaction as they arise in the treatment of dynamical screening or phonons, or in the context of extended dynamical mean-field theory.
It also opens the way to the numerical implementation of the recently proposed dual boson approach and thereby to the calculation of properties of models with long-range interaction.
The computation of the vertex functions may further be employed to obtain momentum resolved response functions in the context of dynamical mean-field theory.
Within dual fermion calculations, one can use this solver to include the effects of dynamical screening while accounting for dynamical spatial correlations at the same time. This will be relevant for an accurate description of real materials.

Results for the self-energy and vertex functions have been obtained within dynamical mean-field theory including the effects of a retarded interaction.
While in the antiadiabatic regime the self-energy is governed by a low-energy effective model at small energies and approximated by the so-called dynamic atomic limit approximation at high frequencies, the present solver provides an efficient and unbiased method for a general retarded interaction.
The three-leg vertex function has been seen to be strongly enhanced when approaching the Mott transition. Because of the fact that it develops significant structure one may expect vertex corrections in extensions of extended dynamical mean-field theory to be important, in particular close to the Mott transition.
The structures in the three-leg and four-leg vertex functions in the vicinity of the Mott transition are found to be generic features of the transition, which are not affected by the particular choice of screening model or how the transition is approached.

An implementation of the improved estimators for the self-energy and vertex function for multi-orbital impurity models with retarded interaction is provided as an open source code~\cite{Hafermann13}, as part of the ALPS libraries~\cite{ALPS2}.

\acknowledgments
The author would like to thank Thomas Ayral, Silke Biermann, Alexander Lichtenstein, Erik van Loon, Junya Otsuki, Olivier Parcollet and Philipp Werner for helpful  discussions and comments. The simulations have been performed using an implementation based on the ALPS-libraries~\cite{ALPS2}. The author acknowledges support from the FP7/ERC, under Grant Agreement No. 278472-MottMetals.


\appendix

\section{Impurity action}
\label{app:impurity}

The impurity model action may be decomposed into the following three parts:

\begin{align}
S =  S_{\text{at}} + S_{\text{Fermion}} + S_{\text{Boson}}
\end{align}
with
\begin{align*}
S_{\text{at}} = & -\int_{0}^{\beta}d\tau\int_{0}^{\beta}d\tau'\sum_{ij}c_{i}^{*}(\tau)\mathcal{G}^{0\, -1}_{ij}(\tau-\tau')c_{j}(\tau') \notag\\
&+ \frac{1}{2}\sum_{ij}U_{ij}\int_{0}^{\beta}\!\!\!d\tau\, n_{i}(\tau)n_{j}(\tau)\\
S_{\text{Fermion}} =& -\int_{0}^{\beta}d\tau\int_{0}^{\beta}d\tau'\sum_{\kv ij}f^{*}_{\kv i}(\tau)G^{f\,-1}_{\kv ij}(\tau-\tau') f_{\kv j}(\tau')\notag\\
& + \int_{0}^{\beta}d\tau \sum_{\kv ij} \left[c_{i}^{*}(\tau)V_{\kv ij}f_{\kv j}(\tau)   + f_{\kv i}^{*}(\tau)V^{*}_{\kv ij}c_{j}(\tau) \right]\\
S_{\text{Boson}} =& -\int_{0}^{\beta}d\tau\int_{0}^{\beta}d\tau'\sum_{\qv} b_{\qv}^{*}(\tau)\tilde{\mathcal{D}}^{-1}_{\qv}(\tau-\tau')b_{\qv}(\tau') 
\notag\\
&+ \int_{0}^{\beta}d\tau\sum_{\qv i}[b^{*}_{\qv}(\tau)+b_{\qv}(\tau)]\lambda_{\qv}n_{i}(\tau).
\end{align*}
In Fourier representation, the different propagators read
\begin{align}
G^{f}_{\kv ij}(\inu_{n}) &= [\inu_{n}-\epsilon_{\kv}^{i}]^{-1}\delta_{ij},\nonumber\\
\mathcal{G}^{0}_{ij}(\inu_{n}) &= [\inu_{n}+\mu-\epsilon_{i}]^{-1}\delta_{ij},\nonumber\\
\tilde{\mathcal{D}}_{\qv ij}(\iom_{m})&= [\iom_{m}-\omega_{\qv}]^{-1},
\end{align}
where $\nu_{n}=(2n+1)\pi/\beta$ is a fermionic Matsubara frequency and $\omega_{m}=2m\pi/\beta$ is bosonic.

The fermionic degrees of freedom can be integrated out using the following identity for Grassmann variables,
\begin{align}
\label{hstfermion2}
&\int \prod_{k}df_{k}^{*}df_{k} e^{-f_{i}^{*}H_{ij}f_{j} + c^{*}_{i}b_{ij}f_{j}  + f^{*}_{i}b^{*}_{ij}c_{j} }\notag\\
&= [\det H] e^{c^{*}_{i}[bH^{-1}b^{*}]_{ij}c_{j}},
\end{align}
which gives rise to the hybridization function
\begin{align}
\sum_{\kv\, kl}V_{\kv ik}G_{\kv kl}(\inu)V_{\kv lj}^{*} = \sum_{\kv\,l}\frac{V_{\kv il}V_{\kv lj}^{*}}{\inu-\epsilon_{\kv l}} \teL \Delta_{ij}(\inu).
\end{align}
In complete analogy one may integrate out the free bosons using the identity
\begin{align}
\label{hstboson}
\int \prod_{k} \frac{db^{*}_{k} db_{k}}{2\pi i} e^{-b_{i}^{*}H_{ij}b_{j} + J_{i}^{*}b_{i} + J_{i}b_{i}^{*} } = [\det H]^{-1} e^{J_{i}^{*}H^{-1}_{ij}J_{j}}.
\end{align}
This leads to
\begin{align}
S_{\text{ret}}&=\sum_{\qv\,ij}\int_{0}^{\beta}d\tau\int_{0}^{\beta}d\tau' n_{i}(\tau)\lambda_{\qv}\tilde{\mathcal{D}}_{\qv}(\tau-\tau) \lambda_{\qv}n_{j}(\tau'),
\end{align}
where the noninteracting propagator of the field $b$ is defined as
\begin{align}
\tilde{\mathcal{D}}_{\qv}(\tau-\tau') \Let - \av{b_{\qv}(\tau)b_{\qv}^{*}(\tau')}_{0},
\end{align}
which is complex. It is more convenient to work with a propagator which is real, since in particular the retarded interaction is real.
This is accomplished by considering the propagator of the real field $b+b^{*}$, for which
\begin{align}
\mathcal{D}_{\qv}(\iom) = \tilde{\mathcal{D}}_{\qv}(\iom) + \tilde{\mathcal{D}}^{*}_{\qv}(\iom)
\end{align}
holds.
Note that when written in terms of this propagator, the retarded part of action carries a factor $1/2$:
\begin{align}
\label{sretd}
S_{\text{ret}}&=\frac{1}{2}\sum_{\qv\,ij}\int_{0}^{\beta}d\tau\int_{0}^{\beta}d\tau' n_{i}(\tau)\lambda_{\qv}\mathcal{D}_{\qv}(\tau-\tau) \lambda_{\qv}n_{j}(\tau').
\end{align}
In imaginary time, the bosonic propagator reads
\begin{align}
\label{dtau}
\mathcal{D}_{\qv}(\tau) = \tilde{\mathcal{D}}_{\qv}(\tau)  + \tilde{\mathcal{D}}_{\qv}(\beta-\tau)
&=  -\frac{e^{\omega_{\qv}\tau}}{e^{\omega_{\qv}\beta}-1} - \frac{e^{-\omega_{\qv}\tau}}{1-e^{-\omega_{\qv}\beta}} \notag\\
&=-\frac{\cosh[(\tau-\beta/2)\omega_{\qv}]}{\sinh(\omega_{\qv}\beta/2)}.
\end{align}
Defining the retarded interaction as
\begin{align}
\label{uretdef}
U_{\text{ret}}(\tau-\tau') \Let \sum_{\qv} \lambda_{\qv} \mathcal{D}_{\qv}(\tau-\tau') \lambda_{\qv}
\end{align}
yields the result \eqref{sret}.
By introducing the operators and corresponding conjugate momenta
\begin{align}
\phi_{\qv}\Let \frac{1}{\sqrt{2}}(b_{\qv}^{\dagger}+ b_{\qv}),\qquad \Pi_{\qv}\Let \frac{1}{i\sqrt{2}}(b_{\qv}^{\dagger} - b_{\qv})
\end{align}
which obey $[\phi_{\qv},\Pi_{\qv'}] = i\delta_{\qv\qv'}$, one obtains an alternative representation for the bosonic part of the Hamiltonian, Eqs. \eqref{hboson1} and \eqref{hboson2}. Up to an irrelevant additive constant it can be rewritten in the following form:
\begin{align}
H_{\text{Boson}} =\sum_{\qv}\frac{\omega_{\qv}}{2}\left(\phi_{\qv}^{2} + \Pi_{\qv}^{2}\right) + \sum_{\qv}\sqrt{2}\phi_{\qv}\lambda_{\qv}\sum_{i}n_{i}.
\end{align}
Using the equation of motion for the Heisenberg operators $b_{\qv}^{(\dagger)}(\tau)$, one finds that
\begin{align}
[\partial_{\tau}\phi_{\qv}(\tau)]^{2} = - \omega_{\qv}^{2}\Pi_{\qv}^{2}(\tau),
\end{align}
or $\Pi_{\qv}^{2}(\iom) = -(\iom)^{2}\phi_{\qv}^{2}/\omega_{\qv}^{2}$. Passing to the action formulation, one therefore obtains
\begin{align}
\label{sbosonphi}
S_{\text{Boson}}
=&\sum_{m}\sum_{\qv}\phi_{\qv}(\iom_{m})\left[\frac{-(\iom_{m})^{2}+\omega_{\qv}^{2}}{2\omega_{\qv}}\right]\phi_{\qv}(-\iom_{m})\notag\\
&+ \sum_{m}\sum_{\qv i} \sqrt{2}\phi(\iom_{m})\lambda_{\qv}n_{i}(-\iom_{m})\notag\\
=&-\frac{1}{2}\sum_{m}\sum_{\qv}\tilde{\phi}_{\qv}(\iom_{m})\mathcal{D}^{-1}_{\qv}(\iom_{m})\tilde{\phi}_{\qv}(-\iom_{m})\notag\\
&+ \sum_{m}\sum_{\qv i} \tilde{\phi}_{\qv}(\iom_{m})\lambda_{\qv}n_{i}(-\iom_{m}),
\end{align}
where in the second line the rescaled fields $\tilde{\phi}=\sqrt{2}\phi$ have been introduced for convenience. The propagator in \eqref{sbosonphi} is the same as in \eqref{sretd}, since $\tilde{\phi}=b^{*}+b$.
One may easily verify that both formulations lead to the same results. In particular, one may use the identity
\begin{align}
\label{realhst}
\int \frac{\prod_{i} d\phi_{i}}{\sqrt{(2\pi)^{N} \det W}} e^{-\frac{1}{2}\phi_{i}W^{-1}_{ij}\phi_{j} \pm \phi_{i} n_{i}}
= e^{\frac{1}{2}n_{i}W_{ij}n_{j}}
\end{align}
to integrate out the $\tilde{\phi}$-fields, which recovers \eqref{sretd}.

\subsection{Generating function}

For the following derivation, one needs the generating function for correlation functions involving bosonic fields.
Introducing sources and integrating out the fields using \eqref{realhst} yields
\begin{align}
\label{genfunc}
&G_{\text{Boson}}[J] = \int\mathcal{D}[\tilde{\phi}] e^{-S_{\text{Boson}}[\tilde{\phi}] + \sum_{\qv} J_{\qv}(\tau)*\tilde{\phi}_{\qv}(\tau)} \notag \\
 &= e^{\frac{1}{2}\sum_{\qv}\sum_{ij}(J_{\qv}(\tau)-\lambda_{\qv}n_{i}(\tau))*\mathcal{D}_{\qv}(\tau-\tau')*(J_{\qv}(\tau')-\lambda_{\qv}n_{j}(\tau'))}
\end{align}
where the '$*$' denotes time integration.
Integrating out a single field $\tilde{\phi}_{\qv}(\tau)$ from a given expression '$\ldots$' is hence accomplished by taking the corresponding functional derivative of the generating function:
\begin{align}
\label{intphi}
\int \mathcal{D}[\tilde{\phi_{\qv}}] \tilde{\phi_{\qv}}(\tau)\ldots =& \left. \frac{\delta G_{\text{Boson}}[J]}{\delta J_{\qv}(\tau)}\right|_{J=0}\ldots\notag\\
=& \sum_{i} \int_{0}^{\beta} d\tilde{\tau} n_{i}(\tilde{\tau})\mathcal{D}_{\qv}(\tilde{\tau}-\tau)\lambda_{\qv} \ldots
\end{align}

\subsection{Improved estimators}
\label{app:impest}

The derivation is similar to the one in Ref.~\onlinecite{Hafermann12}. The difference is that the commutator with the Hamiltonian \eqref{hamiltonian} in Eq. \eqref{eom} generates an additional term due to the coupling to a bosonic bath:
\begin{align}
\label{hcomm}
[H, c_a] = -\varepsilon_{a} c_a - \sum_{\kv j} V_{\kv}^{aj} f_{\kv j}  &-\sum_j n_{j}U_{ja} c_a \notag\\
&- \sum_{\qv}(b^{\dagger}_{\qv}+ b_{\qv})\lambda_{\qv}c_{a}.
\end{align}
This extra term gives rise to an additional correlation function
\begin{align}
F^{\text{ret}}_{ab}(\tau-\tau') = -\sum_{\qv}\lambda_{\qv}\av{T_{\tau} \tilde{\phi}_{\qv}(\tau) c_{a}(\tau)c_{b}^{\dagger}(\tau')},
\end{align}
where $\tilde{\phi}_{\qv}$ has been substituted for $b_{\qv}^{\dagger} + b_{\qv}$. 
The correlation function can be further evaluated by switching to the path integral representation and using Eq.~\eqref{intphi} to integrate out field $\tilde{\phi}_{\qv}$. With the definition \eqref{uretdef} of the retarded interaction, one obtains the final expression \eqref{fretfinal}.

Similarly, the equation of motion generates an extra term for the improved estimator of the vertex function:
\begin{align}
&F^{(4),\text{ret}}_{abcd}(\tau_{a},\tau_{b},\tau_{c},\tau_{d})\notag\\
&=\sum_{\qv}\lambda_{\qv}\av{T_{\tau} \tilde{\phi}_{\qv}(\tau_{a}) c_{a}(\tau_{a})c_{b}^{\dagger}(\tau_{b})c_{c}(\tau_{c})c_{d}^{\dagger}(\tau_{d})}.
\end{align}
Integrating out the boson fields as before results in \eqref{hretfinal}.

For the three-leg vertex one uses the following identity
\begin{align}
\partial_{\tau_{a}}\!\!\av{T_{\tau} c_{a}(\tau_{a})c_{b}^{\dagger}(\tau_{b})n_{c}(\tau_{c})}
=&
\av{T_{\tau} \partial_{\tau_{a}} c_{a}(\tau_{a})c_{b}^{\dagger}(\tau_{b})n_{c}(\tau_{c})}\notag\\
&+\delta(\tau_{a}-\tau_{b})\delta_{ab}\av{n_{c}}\notag\\
&+\delta(\tau_{a}-\tau_{c})\delta_{ac}\notag\\
&\,\times\av{T_{\tau}c_{c}(\tau_{c})c_{b}^{\dagger}(\tau_{b})}.
\end{align}
The delta-function contributions stem from the discontinuities of this function. The rest of the derivation follows Ref.~\onlinecite{Hafermann12}. Inserting the commutator \eqref{hcomm}
yields
\begin{align}
&-[\partial_{\tau_{a}}+\epsilon_{a}]G^{(3)}_{abc}(\tau_{a},\tau_{b},\tau_{c})
=
G^{(3),fc}_{abc} (\tau_{a},\tau_{b},\tau_{c})
\notag\\
&+ F^{(3)}_{abc}(\tau_{a},\tau_{b},\tau_{c})+F^{(3),bc}_{abc}(\tau_{a},\tau_{b},\tau_{c})
\notag\\
&+\delta(\tau_{a}-\tau_{b})\delta_{ab}\av{n_{c}}-\delta(\tau_{a}-\tau_{c})\delta_{ac}G_{cb}(\tau_{a}-\tau_{b}),
\end{align}
where the following correlation functions have been defined
\begin{align}
G^{(3)}_{abc}(\tau_{1},\tau_{b},\tau_{c}) \Let & -\av{T_{\tau} c_{a}(\tau_{a})c_{b}^{\dagger}(\tau_{b})n_{c}(\tau_{c})}\notag\\
G^{(3),fc}_{abc} (\tau_{1},\tau_{b},\tau_{c}) \Let & -\sum_{\kv j}V_{\kv}^{aj}\av{T_{\tau} f_{\kv j}(\tau_{a})c_{b}^{\dagger}(\tau_{b})n_{c}(\tau_{c})}\notag\\
F^{(3),\text{st}}_{abc}(\tau_{1},\tau_{b},\tau_{c}) \Let & -\sum_{j}\av{T_{\tau} n_{j}(\tau_{a})U_{ja}c_{a}(\tau_{a})c_{b}^{\dagger}(\tau_{b})n_{c}(\tau_{c})}\notag\\
F^{(3),bc}_{abc}(\tau_{1},\tau_{b},\tau_{c}) \Let & -\sum_{\qv}\lambda_{\qv}\av{T_{\tau}\tilde{\phi}_{\qv}(\tau_{a})c_{a}(\tau_{a})c_{b}^{\dagger}(\tau_{b})n_{c}(\tau_{c})}.
\end{align}
The last correlation function is evaluated in analogy to the foregoing (Appendix~ \ref{app:impest}):
\begin{align}
&F^{(3),\text{ret}}_{abc}(\tau_{1},\tau_{b},\tau_{c}) \Let  \notag\\
&-\int_{0}^{\beta}d\tilde{\tau}\sum_{i}\av{n_{i}(\tilde{\tau})U_{\text{ret}}(\tilde{\tau}-\tau_{a}) c_{a}(\tau_{a})c_{b}^{*}(\tau_{b})n_{c}(\tau_{c})}.
\end{align}
Taking the Fourier transform
\begin{align}
&f(\inu,\iom) = \mathcal{F}[f(\tau_{a},\tau_{b},\tau_{c})]\Let\notag\\
&\frac{1}{\beta}\int_0^\beta d\tau_a \int_0^\beta d\tau_b \int_0^\beta d\tau_c f(\tau_{a},\tau_{b},\tau_{c})e^{\inu\tau_a}e^{-\i(\nu+\omega)\tau_b}e^{\iom\tau_c}
\end{align}
and expressing $G_{abc}^{(3),fc}$ in terms of $G^{(3)}_{abc}$ through its equation of motion yields
\begin{align}
&\sum_{j}[(\inu-\epsilon_{a})\delta_{aj}-\Delta_{aj}(\inu)] G^{(3)}_{jbc} (\inu,\iom)\notag\\
&=
\ F^{(3)}_{abc}(\inu,\iom)+\beta\delta_{ab}\av{n_{c}}\delta_{\omega}-\delta_{ac}G_{cb}(\inu+\iom).
\end{align}
Subtracting $\sum_{j}\Sigma_{aj}(\inu)G^{(3)}_{jbc} (\inu,\iom)$ on both sides, the left-hand side becomes $\sum_{j}G^{-1}_{aj}(\inu)G^{(3)}_{jbc} (\inu,\iom)$. Hence multiplying both sides by $G$ and defining the connected part 
\begin{align}
G^{(3),\text{con}}_{abc}(\inu,\iom) = G^{(3)}_{abc}(\inu,\iom) -[&\beta G_{ab}(\inu)\av{n_{c}}\delta_{\omega,0}\notag\\
&-G_{ac}(\inu)G_{cb}(\inu+\iom)],
\end{align}
as well as using $F'\Let G\Sigma$, one finally obtains
\begin{align}
G^{(3),\text{con}}_{abc} (\inu,\iom) 
=& \sum_{i} G_{ai}(i\nu)F^{(3)}_{ibc}(\inu,\iom)
\notag\\
&-\sum_{i}F'_{ai}(\inu)G^{(3)}_{ibc} (\inu,\iom).
\end{align}

\section{Self-energy tails}
\label{app:tails}

We are interested in the high-frequency behavior of the self-energy up to first order in $1/(\inu_{n})$,
\begin{align}
\label{sigmaexp}
\Sigma(\inu) = \Sigma^{0} + \frac{\Sigma^{1}}{\inu} + \ldots
\end{align}
Expanding
\begin{align}
\label{grelsigm}
G(\inu) = \left[(\inu)\um-\hat{\varepsilon}-\Delta(\inu)-\Sigma(\inu)\right]^{-1}
\end{align}
in $1/\inu$, using \eqref{sigmaexp} and the corresponding expansion for $\Delta(\inu)$ up to first order,
\begin{align}
\Delta_{ab}(\inu) = \sum_{\kv i}\frac{V_{\kv}^{ai}V_{\kv}^{*\,ib}}{\inu-\epsilon_{\kv}} =\frac{\Delta_{ab}^{1}}{\inu} + \ldots
\end{align}
one obtains the following expression in matrix form:
\begin{align}
\label{gexp}
G(\inu) &= \frac{\um}{\inu} + \frac{\hat{\varepsilon}+\Sigma^{0}}{(\inu)^{2}} + \frac{({\hat{\varepsilon}}+\Sigma^{0}_{a})(\hat{\varepsilon}+\Sigma^{0})+\Delta^{1}+\Sigma^{1}}{(\inu)^{3}} + \ldots
\end{align}
where $\hat{\varepsilon}_{ab}=(\varepsilon_{a}-\mu)\delta_{ab}$.
The high-frequency expansion of the Green's function is computed from the well-known expression
\begin{align}
\label{gexp2}
G_{ab}(\inu) = \sum_{k=0}^{\infty} (-1)^{k}\frac{\av{\{[H,c_{a}]_{\{k\}},c_{b}^{\dagger} \}}}{(\inu)^{k+1}},
\end{align}
where $[H,c_{a}]_{\{k\}}$ denotes the $k$-fold nested commutator with the Hamiltonian and $\{A,B\}$ denotes the anticommutator of $A$ and $B$.
Using \eqref{hcomm},  defining $C^{1}_{a}$ and $C^{2}_{a}$ such that
$\langle\{[H, c_a],c_{b}^{\dagger}\}\rangle =C^{1}_{a}\delta_{ab}$ and $\langle \{[H,[H,c_{a}]],c_{b}^{\dagger}\} \rangle = \Delta^{1} + C^{2}_{a}\delta_{ab}$, one finds
\begin{align}
\label{c1}
C^{1}_{a} = -\varepsilon_{a} -\sum_j \langle  n_j \rangle U_{ja}
- \sum_{\qv}\lambda_{\qv}\langle\tilde{\phi}_{\qv}\rangle
\end{align}
and
\begin{align}
\label{c2}
C^{2}_{a} =& \epsilon_{a}^{2} + 2\sum_{j}\epsilon_{a}U_{aj}\langle n_{j}\rangle + \sum_{ij}U_{ai}U_{aj}\langle n_{i}n_{j}\rangle\notag\\
&+2\sum_{\qv j}U_{aj}\lambda_{\qv}\langle \tilde{\phi}_{\qv}n_{j}\rangle+ 2\sum_{\qv}\epsilon_{a}\lambda_{\qv}\langle \tilde{\phi}_{\qv}\rangle\notag\\
&-\i\sum_{\qv}\omega_{\qv}\lambda_{\qv}\langle \tilde{\Pi}_{\qv}\rangle
+\sum_{\qv\qv'}\lambda_{\qv}\lambda_{\qv'} \langle \tilde{\phi}_{\qv}\tilde{\phi}_{\qv'}
 \rangle.
\end{align}
Comparing \eqref{gexp} with \eqref{gexp2}, one arrives at
\begin{align}
\label{sigma_tail0interm}
\Sigma_{a}^{0} = \sum_{j}U_{aj}\langle n_{j}\rangle + \sum_{\qv}\lambda_{\qv}\langle \tilde{\phi}_{\qv}\rangle
\end{align}
and
\begin{align}
\label{sigma_tail1interm}
\Sigma_{a}^{1} = -\i\sum_{\qv}\omega_{\qv}\lambda_{\qv}\langle\tilde{\Pi}_{\qv}\rangle& + \sum_{ij}U_{ai}U_{aj}\Big(\langle n_{i}n_{j}\rangle - \langle n_{i}\rangle\langle n_{j}\rangle\Big)\notag\\
& + 2 \sum_{\qv\,j}U_{aj}\lambda_{\qv}\Big(\langle\tilde{\phi}_{\qv} n_{j}\rangle - \langle\tilde{\phi}_{\qv}\rangle\langle n_{j}\rangle\Big)\notag\\
& + \sum_{\qv\qv'}\lambda_{\qv}\lambda_{\qv'}\Big( \langle \tilde{\phi}_{\qv}\tilde{\phi}_{\qv'}\rangle - \langle \tilde{\phi}_{\qv}\rangle\langle\tilde{\phi}_{\qv'}\rangle \Big). 
\end{align}
In order to further evaluate this expression, one uses the generating function, Eq. \eqref{intphi}, to integrate out the fields $\tilde{\phi_{\qv}}$ and to compute the correlation functions
\begin{align}
\label{phiav}
\langle \tilde{\phi}_{\qv}(\tau)\rangle &= \sum_{i}\av{n_{i}} \int_{0}^{\beta} d\tau' D_{\qv}(\tau-\tau')\lambda_{\qv},\\
\langle\tilde{\phi}_{\qv}(\tau)n_{j}(0)\rangle &=
\sum_{i} \int_{0}^{\beta} d\tau' D_{\qv}(\tau-\tau')\av{n_{i}(\tau')n_{j}(0)}\lambda_{\qv} 
\end{align}
as well as the interacting Boson propagator
\begin{align}
D_{\qv\qv'}(\tau-\tau') \Let -\langle\tilde{\phi}_{\qv}(\tau)\tilde{\phi}_{\qv'}(\tau')\rangle,
\end{align}
by taking the second derivative of the generating function \eqref{genfunc} with respect to $J$:
\begin{align}
\label{phiphiav}
& D_{\qv\qv'}(\tau-\tau') = \mathcal{D}_{\qv}(\tau-\tau')\delta_{\qv\qv'} -\int_{0}^{\beta}\!\!d\tau_{1} \int_{0}^{\beta}\!\!d\tau_{2}
\sum_{ij}\notag\\ & \times\lambda_{\qv} \mathcal{D}_{\qv}(\tau-\tau_{1})\av{n_{i}(\tau_{1})n_{j}(\tau_{2})} \mathcal{D}_{\qv'}(\tau_{2}-\tau')\lambda_{\qv'}.
\end{align}
The expectation values in \eqref{c1}, \eqref{c2} involving $\tilde{\phi}_{\qv}$ can now be expressed in terms of time ordered correlation functions
\begin{align}
\langle \tilde{\phi}_{\qv}\rangle &\equiv \langle \tilde{\phi}_{\qv}(0^{+})\rangle,\\
\langle\tilde{\phi}_{\qv}n_{j}\rangle &= \langle\tilde{\phi}_{\qv}(0^{+})n_{j}(0)\rangle,\\
\langle\tilde{\phi}_{\qv}\tilde{\phi}_{\qv'}\rangle &= \langle\tilde{\phi}_{\qv}(0^{+})\tilde{\phi}_{\qv'}(0)\rangle.
\end{align}
Noting that $\langle\Pi_{\qv}\rangle=0$ and substituting \eqref{phiav}-\eqref{phiphiav} into \eqref{sigma_tail0interm}, \eqref{sigma_tail1interm} and using \eqref{uretdef}, yields the final expressions
\begin{align}
\Sigma_{a}^{0} =&  \sum_{j} U_{aj}\av{n_{j}} + \sum_{i}\av{n_{i}}\int_{0}^{\beta} d\tau U_{\text{ret}}(\tau),\\
\Sigma_{a}^{1} =& -U_{\text{ret}}(0^{+}) + \sum_{ij} U_{ai} U_{aj} \Big(\av{n_{i}n_{j}}-\av{n_{i}}\av{n_{j}}\Big)\notag\\
&+ 2\sum_{ij} U_{aj} \int_{0}^{\beta}d\tau U_{\text{ret}}(\tau)\Big(\av{n_{i}(\tau)n_{j}(0)}-\av{n_{i}}\av{n_{j}}\Big)\notag\\
&+\sum_{ij} \int_{0}^{\beta}d\tau \int_{0}^{\beta}d\tau' U_{\text{ret}}(\tau)U_{\text{ret}}(\tau')\times\notag\\
&\qquad\Big(\av{n_{i}(\tau)n_{j}(\tau')}-\av{n_{i}}\av{n_{j}}\Big),\end{align}
which can be brought into the compact form \eqref{sigma_tail}-\eqref{sigma_tail1}
in terms of the retarded interaction $U_{ij}(\tau)=U_{ij}\delta(\tau) + U_{\text{ret}}(\tau)$.

\section{Local bosonic propagator}
\label{app:polarization}

In terms of the fields
\begin{align}
\tilde{\phi}\Let\sum_{\qv}\lambda_{\qv}\tilde{\phi}_{\qv}
\end{align}
one can rewrite the bosonic part of the action \eqref{sbosonphi} in the form
\begin{align}
\label{sbosonphiloc}
S_{\text{Boson}}
=&-\frac{1}{2}\sum_{m}\tilde{\phi}(\iom_{m})\mathcal{D}^{-1}(\iom_{m})\tilde{\phi}(-\iom_{m})\notag\\
&+ \sum_{m}\sum_{i} \tilde{\phi}(\iom_{m})n_{i}(-\iom_{m}).
\end{align}
The local bosonic propagator of the impurity model is then defined as follows:
\begin{align}
D(\tau-\tau')=-\av{\tilde{\phi}(\tau)\tilde{\phi}(\tau')}.
\end{align}
In terms of the propagator of the bosonic bath it can be expressed as
\begin{align}
D(\tau-\tau') = \sum_{\qv\qv'}\lambda_{\qv}D_{\qv\qv'}(\tau-\tau')\lambda_{\qv'}.
\end{align}
From the corresponding relation for the bare propagators, one sees that the retarded interaction \eqref{uretdef} plays the role of the bare local bosonic propagator $\mathcal{D}=-\langle\tilde{\phi}(\tau)\tilde{\phi}(\tau')\rangle_{0}$.
Using \eqref{phiphiav}, the bosonic propagator can be written
\begin{align}
\label{bosonicp}
D(\tau-\tau') =& \mathcal{D}(\tau-\tau') -\int_{0}^{\beta}\!\!d\tau_{1} \int_{0}^{\beta}\!\!d\tau_{2}\mathcal{D}(\tau-\tau_{1}) \notag\\
& \qquad\times\av{n(\tau_{1})n(\tau_{2})} \mathcal{D}(\tau_{2}-\tau'),
\end{align}
where $n=\sum_{i}n_{i}$ is the total charge density. When the retarded part of the action, Eq. \eqref{sretd}, is written in terms of $n-\av{n}$ instead of $n$ itself, the average in the above equation becomes $\av{\bar{n}(\tau_{1})\bar{n}(\tau_{2})}$. In this case, the above expression can be written in terms of the local charge susceptibility
\begin{align}
\chi(\tau_{1}-\tau_{2}) = -\sum_{ij}\av{\bar{n}_{i}(\tau_{1})\bar{n}_{j}(\tau_{2})}
\end{align}
on Matsubara frequencies as
\begin{align}
& D(\iom) = \mathcal{D}(\iom) + \mathcal{D}(\iom)\chi(\iom) \mathcal{D}(\iom).
\end{align}
From this relation the bosonic self-energy (impurity polarization) $\Pi=\mathcal{D}^{-1}-D^{-1}$ is identified to be
\begin{align}
\Pi(\iom) = \frac{\chi(\iom)\mathcal{D}(\iom)}{D(\iom)}.
\end{align}

\bibliography{main}

\end{document}